\documentclass[12pt,preprint]{aastex}

\usepackage{graphicx}
\usepackage{rotating}

\shorttitle{}
\shortauthors{Nesvorn\'y et al.}

\begin{document}
\baselineskip 19.pt

\title{Terrestrial Planet Formation from Two Source Reservoirs}

\author{David Nesvorn\'y$^1$, Alessandro Morbidelli$^2$, William F. Bottke$^1$, \\Rogerio Deienno$^1$,
Max Goldberg$^2$}

\affil{(1) Solar System Science \& Exploration Division, Southwest Research Institute, 1301 Walnut St., 
  Suite 400,  Boulder, CO 80302, USA}

\affil{(2) Laboratoire Lagrange, UMR7293, Universit\'e C\^ote d'Azur, CNRS, Observatoire de la 
C\^ote d'Azur, Bouldervard de l'Observatoire, 06304, Nice Cedex 4, France}

\begin{abstract}
  This work describes new dynamical simulations of terrestrial planet formation. The simulations started
  at the protoplanetary disk stage, when planetesimals formed and accreted into protoplanets, and continued
  past the late stage of giant impacts. We explored the effect of different parameters, such as the initial
  radial distribution of planetesimals and Type-I migration of protoplanets, on the final results. In
  each case, a thousand simulations were completed to characterize the stochastic nature of the accretion process.
  In the model best able to satisfy various constraints, Mercury, Venus, and Earth accreted from planetesimals 
  that formed early near the silicate sublimation line at $\simeq 0.5$ au and migrated by disk torques. For Venus
  and Earth to end up at 0.7-1 au, Type-I migration had to be directed outward, for example as the 
  magnetically driven winds reduced the surface gas density in the inner part of the disk. Mercury was left behind
  near the original ring location. We suggest that Mars and multiple Mars-sized protoplanets grew from a distinct
  outer source of planetesimals at 1.5-2 au. While many migrated inwards to accrete onto the proto-Earth, our
  Mars was the lone survivor. This model explains: (1) the masses and orbits of the terrestrial planets,
  (2) the chemical composition of the Earth, where $\sim 70$\% and $\sim 30$\% come from reduced inner-ring
  and more-oxidized outer-ring materials, and (3) the isotopic differences of the Earth and Mars. It suggests
  that the Moon-forming impactor Theia plausibly shared a similar isotopic composition and accretion history
  with that of the proto-Earth.  
\end{abstract}

\section{Introduction}

A problem with the classical model of terrestrial planet accretion from a radially extended disk of planetesimals
(0.3-2~au) is that it fails to reproduce the radial mass concentration of the terrestrial planets (Chambers \&
Wetherill 1999, Agnor et al. 1999, Chambers 2001). 
The final masses of Mercury and Mars obtained in this model are too large (by a factor of $\sim 2$-10; Chambers 2001), 
and Venus and Earth end up radially more separated from each other than the real planets.\footnote{By a factor of 
$\sim 1.4$-2.8 in terms of the radial mass concentration (Chambers 2001; see Eq. (10) here).}
This may suggest that the
initial distribution of planetesimals, from which the terrestrial planets formed, was concentrated near the
orbital radii $r=0.7$-1 au. Indeed, in the accretion model starting from a narrow annulus of planetesimals at
$0.7<r<1$~au (Hansen 2009), the two most massive planets, Venus and Earth, form at $r \simeq 0.7$-1 au, as
needed. In addition, it is often the case that a Mars-mass planet rapidly accretes in the annulus and -- before
it accumulates more mass -- is scattered to $r \simeq 1.5$ au. This model could potentially resolve the
small Mars problem but the origin of such a narrow planetesimal annulus is unclear.

A proposed solution to effectively recreate the initial conditions of Hansen (2009) was the so-called ``Grand Tack''
model, broadly characterized as Jupiter's inward and then outward migration within a protoplanetary gas disk
(Walsh et al. 2011). Numerical simulations showed that this behavior could truncate the planetesimal disk at 1 au.
This evolution history would also produce radial mixing of inner and outer Solar System materials and could
explain the compositional differences observed across the asteroid belt (Gradie \& Tedesco 1982).
A potential issue with this evolution history, however, is that the outward migration of Jupiter (and Saturn) is
difficult to sustain in a low viscosity disk, the kind expected to exist within our solar nebula (Griveaud et al.
2024). Additionally, Jupiter's migration cannot explain the inner annulus truncation at $r \simeq 0.7$ au. Consequently,
recent studies have moved away from the idea that the disk was truncated by dynamical perturbations from
the giant planets. Instead, they have explored other means of radially concentrating mass.

Along these lines, a class of recent models considered {\it convergent} migration of planetesimals (Ogihara et
al. 2018a) or protoplanets (Ogihara et al. 2018b; Bro\v{z} et al. 2021) toward $r \simeq 1$ au. For planetesimals
to take on this behavior, they would have to be small enough ($\sim 1$ km) for aerodynamic drag to affect their
orbits, while the surface density of gas would have to peak near 1 au (e.g., perhaps because the inner gas disk
was depleted by magnetically driven winds, or MDWs; e.g., Suzuki et al. 2016). For terrestrial protoplanets
to migrate toward $r \simeq 1$ au, one has invoke disk models in which Type-I migration is directed outward for
$r<1$~au and sunward for $r>1$~au. This can plausibly happen in MDW disks (e.g., Ogihara et al. 2024), 
but other possibilities exist as well (e.g., disks with the inner MRI active and outer dead zones;
Kretke \& Lin 2012, Bitsch et al. 2015, Flock et al. 2017, Jankovic et al. 2022; also see Paardekooper \& Mellema 2006).

Another class of models invoked planetesimal formation in a ring near 1 au. Planetesimal formation could have been
efficient near the silicate and/or water-ice sublimation lines, where various effects concentrate
solids (e.g., Dra{\.z}kowska \& Alibert 2017) and trigger instabilities (Carrera et al. 2021).\footnote{Note
  that the formation of planetesimals in rings near condensation lines does not require the existence
  of pressure bumps at these locations (Morbidelli et al. 2022).}
The silicate line probably started near 1 au in a hot young disk and then rapidly moved to $r \sim 0.5$ au
(Morbidelli et al. 2022). The water-ice line probably started at $\sim 3$-5 au, although some models place
it as close as $r=1.6$ au (Johansen et al. 2021; the location depends on assumed early disk's parameters)
and then moved inward in a cold aging disk. Given the isotopic and chemical composition of the terrestrial
planets and other constraints (Morbidelli et al. 2025), it would be the most straightforward to postulate
their formation from planetesimals near the silicate sublimation line (Morbidelli et al. 2022, Izidoro
et al. 2022, Marschall \& Morbidelli 2023).

In a synthesis model, Woo et al. (2023, 2024) combined planetesimal formation within a ring\footnote{Also see 
Chambers \& Cassen (2002) and Lykawka \& Ito (2017) for similar models with peaked radial distribution of planetesimals.} 
at 1 au with convergent Type-I migration of protoplanets toward 1 au. In this case, both the initial conditions and the
nature of planet migration encouraged tight radial mass concentration of the terrestrial planets.
Their simulations also included the effects of the giant planets undergoing a dynamical instability
(Tsiganis et al. 2005, Nesvorn\'y \& Morbidelli 2012). The net effect of the giant planet instability was
to dynamically excite planetesimals and protoplanets in the terrestrial planet region, which in turn
triggered the final stage of giant impacts (Clement et al. 2018, 2021a; Nesvorn\'y et al. 
2021).\footnote{See Clement et al. (2018) and Woo et al. (2024) for how the timing of the giant planet 
instability affects the terrestrial planet formation. A late instability tends to excessively excite the 
terrestrial planet orbits (also see Agnor \& Lin 2012). An early instability may destabilize protoplanets 
in the Mars region and help a small Mars to form (Clement et al. 2018). The instability also affects the 
timing of the last giant impact on the Earth, which has important implications for the Moon formation 
(Canup et al. 2023).}
By considering various constraints (e.g., the angular momentum deficit of the terrestrial planets;
planetesimal/protoplanetary materials added to the Earth after the Moon-forming impact, with the mass
recorded in the form of highly siderophile elements in the Earth's mantle), Woo et al. (2024)
concluded that the last giant impact more likely occurred relatively early ($\lesssim 80$ Myr after gas
disk dispersal) than late ($\gtrsim 100$ Myr). As for the radial mass concentration (RMC), however, the
results from Woo et al. (2024) were not ideal (see Table 1 in Woo et al. 2024 and Section 4.4 here), even
in the disk models with vigorous Type-I migration toward 1 au.

Here we conducted new dynamical simulations of terrestrial planet growth to understand which combination
of parameters best satisfies constraints (Section 3). We assumed that the birth location of
planetesimals can be described by multi-component models with one or more massive rings and a radially
extended background (Section 2.1). The simulations followed protoplanet accretion during four stages:
(1) gas disk stage (0-5~Myr after the formation of the first solids), (2) giant planet instability
(5-15~Myr), (3) residual giant planet migration (15-50~Myr), and (4) late stage when the giant planets
were already residing on their current orbits (50-300~Myr). We explored a wide range of parameters
(e.g., the mass, orbital location and width of planetesimal rings, direction and strength of Type-I
migration, gas disk lifetime; Section 2). For each parameter choice, we performed a thousand simulations
with slightly modified initial conditions -- generated with different random deviates -- to fully
understand the statistical properties of planetary systems obtained in each case (Section 4). The
implications of this work are discussed in Section 5.  

\section{Model}

Our simulations started with $N$ terrestrial planetesimals that were distributed according to simple analytic 
prescriptions (Section 2.1). The goal was to test different assumptions and see what works.
The simulations employed the {\it Symba} integrator,  an efficient symplectic map (Wisdom \& Holman 1991),
that is capable of accurately modeling encounters between massive bodies (Duncan et al. 1998). We used
$N=400$ to explore different parameter choices, and $N=1000$ or $N=2000$ to test the effect of resolution
in selected cases. The full gravitational interaction between bodies was accounted for in all 
simulations.

The simulations covered four different stages (Sections 2.2-2.4) for a total of 300 Myr after the first solids
had formed (some simulations were terminated at $t=200$ Myr). The bulk of 
our simulations adopted perfect mergers during all impacts between planetesimals/protoplanets, but for selected 
cases we also performed tests with a modified code which accounted for hit-and-run and disruptive 
collisions (Section 2.5). The integrations used a 5~day time step, but we previously verified (Nesvorn\'y et al. 2021) 
that the results were the same when a 3.5 day time step was used (the orbital period of Mercury is about 88 days,
18 and 26 times longer than our time steps. Additional tests with a 3.5 day timestep were performed
here; these tests confirmed the results discussed in Section 4. All simulations were completed on the NASA Pleiades
Supercomputer using a large number of Broadwell (Intel Xeon E5-2680v4) nodes.\footnote{We used about 5 wall-clock
  days on 1000 cores to complete one model. Roughly 100 models were tested. This represents about $10^8$ hours
  of CPU time in total.}

\subsection{Initial conditions}

The probability density function (PDF) for the initial Gaussian-ring distribution of planetesimals is
\begin{equation}
f_1(r)={1 \over \sigma_1 \sqrt{2 \pi} } \exp \left[ -{1 \over 2} \left( {r-r_1 \over \sigma_1} \right)^2 \right] \ ,  
\label{ring}
\end{equation}
where $r_1$ is the mean and $\sigma_1$ is the standard deviation. When the planetesimals were distributed in two
rings (Section 4.8), we used two Gaussian distributions with the ring locations $r_1$ and $r_2$, and the
standard deviations $\sigma_1$ and $\sigma_2$. These distributions were combined as
\begin{equation}
f(r)= (1-w_2) f_1(r) + w_2 f_2(r)\ ,
\label{weight0}
\end{equation}
where $w_2$ was the contribution (weight) of the outer planetesimal reservoir to the total. 

The background planetesimals were distributed in a radially extended disk between $r_{\rm min}$ and $r_{\rm max}$
with a power-law profile. The background PDF is given by 
\begin{equation}
f_3(r)={ \gamma+1 \over {r_{\rm max}^{\gamma+1} - r_{\rm min}^{\gamma+1}} }\ r^\gamma 
\label{back}
\end{equation}
for $\gamma \neq -1$, and 
\begin{equation}
f_3(r)={ 1 \over {\ln r_{\rm max} - \ln r_{\rm min} } }\ r^{-1}
\end{equation}
for $\gamma = -1$. The surface density decreasing as $1/r$, $\Sigma(r) \propto 1/r$, corresponds to $\gamma=0$.

The ring and background populations of planetesimals can be combined together as
\begin{equation}
f(r)= (1-w_3) f_1(r) + w_3 f_3(r)\ ,
\label{weight}
\end{equation}
where $w_3$ was the contribution (weight) of background planetesimals to the total. 
The specific choices of parameters $w_2$, $w_3$, $r_1$, $\sigma_1$, $r_2$, $\sigma_2$,
$r_{\rm min}$, $r_{\rm max}$ and $\gamma$ are discussed
in Section 4. The initial eccentricities and inclinations of planetesimals were distributed according to the 
Rayleigh distribution with the scale parameters $\sigma_e=10^{-4}$ and $\sigma_i=\sigma_e/2$.

In the majority of tested cases, all planetesimal components were added at the beginning of our simulations ($t=0$).
This time roughly corresponds to the start of a protosolar disk’s Class-II stage. In some cases, to test the possibility
that the {\it background} component formed late, the background planetesimals were added at the end of the disk stage, at
$t=5$~Myr in our setup (i.e., at the transition between our Stages 1 and 2; see Sections 2.2 and 2.3). The ring 
planetesimals were always added at $t=0$, except for the models reported in Sections 4.1-4.3, where we ignored Stage 1 and 
tested several previously proposed models without the gas disk.
The total mass of planetesimals was fixed at $M_{\rm tot}=2.1$ $M_{\rm Earth}$ in most of our simulations, 
which approximately produced a correct total mass of final terrestrial planets (Hansen 2009, Woo et al. 2024).
In some simulations, where more mass was wasted (e.g., when small fragments were generated in disruptive collisions
and lost), we examined larger initial masses as well.

All planetesimals and protoplanets were given bulk density $\rho = 3$ g cm$^{-3}$.
The density was kept fixed during all simulations. Note that the real terrestrial planets have
densities 4-5.5 g cm$^{-3}$. This means that large protoplanets in our simulations had slightly enhanced
impact cross-sections compared to the real planets, and this may have modestly decreased the accretion timescale.
To test the importance of this issue, a limited number of simulations were performed with $\rho = 5$ g cm$^{-3}$
(Section 4.3).  

With $N=400$, 1000, and 2000, and the total mass $M_{\rm tot}=2.1$ $M_{\rm Earth}$, our initial `planetesimals'
have the mass $m \simeq 1.5 \times 10^{25}$ g, $6.0 \times 10^{24}$ g and $3 \times 10^{24}$ g, respectively.
For the assumed density of 3 g cm$^{-3}$, the `planetesimal' radii are $r \simeq 1060$ km, 780 km and 620 km,
respectively. This can be compared to Ceres which has the mass $9.1\times10^{23}$ g and radius 473 km.
In reality, the characteristic size of planetesimals arising from the streaming instability is $\sim 100$ km 
(e.g., Klahr \& Schreiber 2020).
This issue highlights a common problem in $N$-body simulations of terrestrial planet formation, namely
that the planetesimal disk is under-resolved. To mimic a real planetesimal disk, the runs would need to
contain the order of $>10^6$ planetesimals.\footnote{The N-body integrator LIPAD is capable of representing
  a large number of planetesimals by a small number of tracer particles (Levison et al. 2012, Walsh \&
  Levison 2016, Deienno et al. 2019). It is difficult to conduct an extensive parametric study with LIPAD
  because the code has very large CPU requirements.} 

There is no simple way around this problem because simulations with $N>10^4$ are prohibitively expensive,
especially when we want to explore a large range of parameters. A superparticle approach is sometimes used
to sidestep this problem, where the behavior of a large number of real bodies is approximated by a small
number of simulated bodies (e.g., Nesvorn\'y et al. 2020, Woo et al. 2024).
We adopted the superparticle approach of Nesvorn\'y et al. (2020) to the problem at hand, tested
it in a few cases, but ultimately did not identify any major differences between the simulations with and
without superparticles.  
This result probably indicates that the properties of the terrestrial planet system were not
established early, by competing processes such as the gravitational self-stirring between planetesimals
and aerodynamics gas drag. They were more fundamentally influenced by the later stages when protoplanets
formed, migrated and experienced giant impacts. For that reason, we opted to not use superparticles in
the simulations reported here.

As for the giant planets, their initial orbits were set to the resonant configuration consistent with
the results of hydrodynamical models of planet-disk interactions in the Type-II regime (e.g., Morbidelli
\& Crida 2007). 
Specifically, Jupiter and Saturn were placed in the 3:2 resonance, with Jupiter at 5.5 au and Saturn at 7.4 au 
(Case 1 in Nesvorn\'y et al. 2013). We did not investigate the 2:1 configuration of Jupiter and Saturn which
would be favored in low-viscosity disks (Griveaud et al. 2024).

The outer planets remained on these orbits during Stage 1 (Section 2.2) -- we assumed that Jupiter/Saturn did not
meaningfully migrate during the gas disk phase (cf. Walsh et al. 2011; see discussion in Section 1).
During the second stage, about 6 Myr after the gas disk dispersal, Jupiter/Saturn were
dynamically kicked out of their resonant orbits by the giant planet instability (Section 2.3). In some
simulations, with the goal to closely reproduce the results of Hansen (2009) and similar setups (Sections 4.1-4.3, we ignored
the gas disk stage and placed Jupiter/Saturn on their current orbits at the beginning of Stage 2. This would
(roughly) be equivalent to a case where the giant planet instability was triggered very early (Liu et al. 2022)
and generated a clean, step-like evolution of the giant planets themselves.  

\subsection{Stage 1: protoplanetary gas disk (0-5 Myr)}

The U-Pb chronometer shows that calcium-aluminum-rich inclusions (CAIs) were the first minerals that formed
in the Solar System, 4.5672 Gyr (Connelly et al. 2012) or 4.5687 Gyr ago (Piralla et al. 2023). This time
is defined as time zero ($t_0$) in cosmochemical chronology (roughly $t=0$ in our simulations). Chondrules
formed in nebular gas in a time period ranging from $\sim 0.7$ to $\sim 4$ Myr after the formation of the
CAIs (e.g., Krot et al. 2005, Budde et al. 2018, Pape et al. 2019, Piralla et al. 2023). Paleomagnetic studies
indicate significant fields in the inner and outer solar system by 3.94 and 4.89 Myr after $t_0$,
respectively, consistent with the nebular gas having dispersed by this time (Weiss et al. 2021).
In our Stage 1 simulations, we therefore schematically assumed that the protoplanetary gas disk lasted 5 Myr
after CAIs. For comparison, the observed protoplanetary disks have typical ages $\sim 2$-10~Myr
(e.g., Haisch et al. 2001, Williams \& Cieza 2011). 

\subsubsection{Disk torques and gas drag}

The simulations of Stage 1 account for: (1) the gas disk torques on protoplanets and (2) aerodynamic gas drag on planetesimals. 
As for (1), planets and planetary embryos in the gas disk were subject to the orbital migration and 
eccentricity/inclination damping as a result of gravitational interaction with gas. The total torque of 
nebular gas on a planet can be expressed as a sum of the Lindblad and corotation torques, $\Gamma=\Gamma_{\rm L}+\Gamma_{\rm C}$ 
(Paardekooper et al. 2010, 2011). It scales with different parameters as 
\begin{equation}
\Gamma = \left ( {q \over h} \right)^2 \Sigma_{\rm g} r^4 \Omega^2\ ,
\end{equation}
where $q$ is the planet-to-star mass ratio ($q=M_{\rm p}/M_*$), $h$ is the disk aspect ratio, $\Sigma_{\rm g}$ is the surface
gas density, and $\Omega$ is the planet's orbital frequency.
We also accounted for the effects of eccentricity and inclination damping 
(Papaloizou \& Larwood 2000, Cresswell \& Nelson 2008). Equations (7-10) in Ogihara et al. (2014) give explicit expressions 
for the migration/damping force vectors that we implemented in the $N$-body codes in this work.

These prescriptions are 
identical to those most recently used in Ogihara et al. (2024). The corotation torque can be positive 
and lead to outward migration of protoplanets if the local gas surface density gradient,
$\partial \ln \Sigma/ \partial \ln r$, is positive (e.g., Paardekooper \& Papaloizou 2009). We included the saturation of the corotation
torque in a situation when the horseshoe libration timescale is shorter than the diffusion timescale (Eqs. (6) and (7) 
in Ogihara et al. 2024). In addition, the corotation torque decreases with increasing eccentricity (Bitsch \& Kley 2010). 
The effective turbulent viscosity of the gas disk was represented by the usual $\alpha$ parameter 
(Shakura \& Sunyaev 1973). We used $\alpha \simeq 10^{-4}$ in most simulations, because recent theoretical and observational
studies favor weak turbulence (e.g., Flaherty et al. 2017; Dullemond et al. 2018), but explored different values as well. 

As for (2), the drag acceleration is given by
\begin{equation}
\mathbf{a}_{\rm d}= - {3 \over 4} {C_{\rm d} \over D} {\rho \over \rho_{\rm p}} v_r \mathbf{v}_r \ ,
\end{equation} 
where $\rho$ is the gas density, $\rho_{\rm p}$ and $D$ are the bulk density and diameter of a planetesimal, 
$\mathbf{v}_r$ is the planetesimal velocity with respect to gas, and $C_{\rm d}$ is the dimensionless aerodynamic drag 
coefficient (Weidenschilling 1977). We implemented the full dependence of the $C_{\rm d}$ parameter on the Mach 
and Reynolds numbers (Brasser et al. 2007). $C_{\rm d}$ approaches the limiting value of 2 in a highly supersonic 
regime.

All initial bodies in our integrations were assigned $D=100$ km, such that they experienced gas drag accelerations
comparable to those felt by newly formed planetesimals (e.g., Klahr \& Schreiber 2020). Once two starting planetesimals
merged with each other, however, the code assigned them diameters that were computed from their mass and
$\rho=3$ g cm$^{-3}$. A smoother transition from planetesimals (affected by gas drag) to protoplanets (affected by
disk torques) is difficult to achieve in our under-resolved planetesimal disk.\footnote{We developed and tested several 
additional methods to deal with the transition from planetesimals to protoplanets. In some simulations, the superparticle 
approach from Nesvorn\'y et al. (2020) was implemented to strictly reproduce the accretion timescale. In other simulations, 
following Woo et al. (2023), we altered the planetesimal mass in the encounter algorithm to correctly approximate 
the effects of dynamical stirring. We found that these different approaches produced similar results. The results
are apparently more sensitive to issues related to the initial distribution of planetesimals, planet migration and
late stage of giant impacts, than to details of early accretion and scattering of planetesimals in the gas disk. Note
that gas produces strong orbital damping, which may help to wipe out any significant dependencies on the treatment of 
early planetesimal accretion/scattering in the simulations.} 
All additional forces discussed here were included in the kick part of the $N$-body integrator.  

\subsubsection{Gas disk profiles}

The gas disk in our simulation is defined by its radial and vertical profiles and its evolution with time. Initially,
we experimented with simple disks. Our surface gas density was given by a power law and the gas was
exponentially removed as
\begin{equation}
\Sigma_{\rm g} = \Sigma_0 \left( { r \over r_0 } \right)^{\beta} \exp (-t/\tau ) \ .
\label{sigma}  
\end{equation}
Here $r_0=1$ au, $\Sigma_0=1700$ g cm$^{-2}$ and $-1.5 \leq \beta \leq -0.5$, as motivated by the concept of the
Minimum Mass Solar Nebula (MMSN; Weidenschilling 1977, Hayashi 1981) and astrophysical disks (Armitage 2011),
and $\tau \sim 1$ Myr (all remaining gas was removed at the end of Stage 1 at $t=5$ Myr). 

These disks did not work for terrestrial planet formation because the strong inward migration carried protoplanets
to $r<0.5$ au (Section 4.4 here; Ogihara et al. 2018b, Woo et al. 2023). We therefore explored a wider range of
possibilities.

Specifically, we examined short-lived disks, disks with low initial gas densities, and disks affected by magnetically 
driven winds (MDWs). The short-lived and/or low-gas-density disks were obtained by reducing $\tau$ and/or
$\Sigma_0$ in Eq. (\ref{sigma}). The MDW disks are affected by stellar irradiation that produces ionization in the 
disk's photosphere. If the electrons produced by the ionization are only weakly coupled with gas atoms, they 
can be carried away by the centrifugal force along curved magnetic lines (if these lines are curved 
enough). This leads to winds that carry away mass and angular momentum, a process that can be especially efficient in 
the inner parts of a disk, within several au of the host star. One important consequence of MDWs is that the surface density 
of gas in the inner disk can have a flat profile or even {\it increase} with radius. If so, $\Sigma(r)$ can evolve 
to a bulging profile with a broad peak at $\sim 1$ au. See examples for $t \sim 1$ Myr old disks in Suzuki et al. 
(2016) and Ogihara et al. (2018a,b, 2024). 

\begin{figure}
\epsscale{0.7}
\plotone{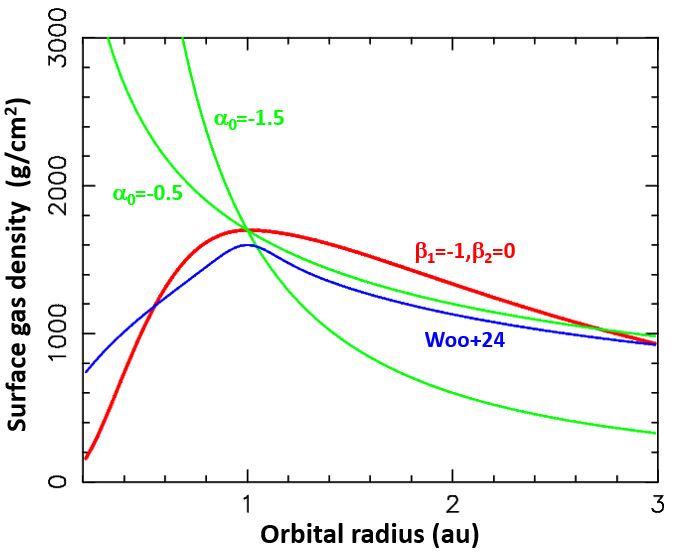}
\caption{The initial surface densities of gas disks considered in this work and elsewhere. The green lines show 
profiles given by Eq. (\ref{sigma}) with $\alpha_0=-0.5$ and $\alpha_0=-1.5$ (MMSN profile), and $\Sigma_0=1700$ g 
cm$^{-2}$. The red line is our MDW disk from Eq. (\ref{sigma2}) with $\Sigma_0=1700$ g cm$^{-2}$, $\beta_1=-1$ and 
$\beta_2=0$. Our MDW disk is reminiscent of disk `shallow' in Woo et al. (2024) (blue line).} 
\label{prof}
\end{figure}

Since the MDW disk profile, and its evolution over time, depends on various unknown parameters (Suzuki et al. 2016,
Ogihara et al. 2018a,b, 2024), here we worked with a simple parameterization of the bulging profile that roughly
approximated the results of MHD simulations and allowed us to test different possibilities. Specifically, 
we used
\begin{equation}
\Sigma_{\rm g} = \Sigma_0 \left( { r \over r_0 } \right)^{\beta(r)} \exp (-t/\tau )   
\label{sigma2}
\end{equation}
with the exponent $\beta(r)=\beta_1 \ln (r/r_0) + \beta_2$. The local slope, $\partial \ln \Sigma / \partial \ln r$, 
is equal to $2 \beta_1 \ln (r/r_0) + \beta_2$ in this case. The $\beta_1$ and $\beta_2$ parameters
define the disk profile. The parameter $\beta_2$ is the local slope at $r=r_0$. Parameter $\beta_1<0$
controls how strongly the slope changes as a function of radius. Figure \ref{prof} shows examples of the
gas disks that we used in this work. 

\begin{figure}
\epsscale{0.8}
\plotone{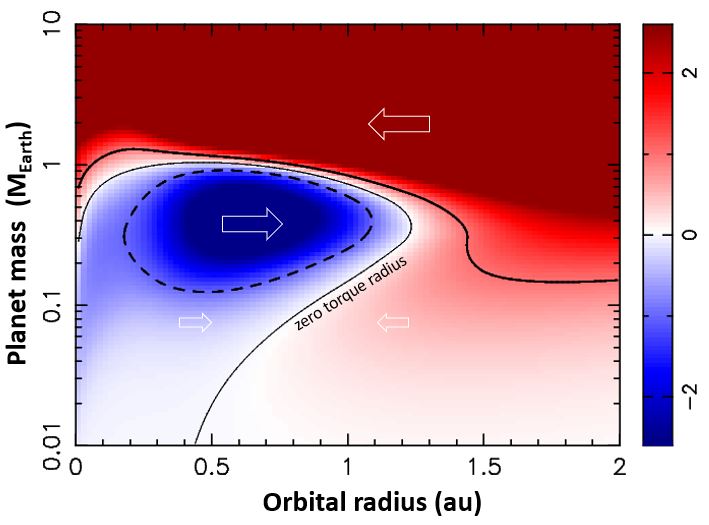}
\caption{A migration map for the MDW disk given by Eq. (\ref{sigma2}) with $\Sigma_0=1700$ g cm$^{-2}$, 
$\beta_1=-1$, $\beta_2=0$, and $\alpha=10^{-4}$. The color scale shows $\delta a/a$, where
$\delta a$ is the change of semimajor axis in 1 Myr. The red and blue colors indicate inward and outward migration,
respectively. The thin solid line is the zero migration radius. The thick solid line is where $-\delta a = a$ 
(strong inward migration) and the dashed line is where $\delta a = a$ (strong outward migration). The migration
speed is given here for $t=0$. It is reduced by the $\exp(-t/\tau)$ factor for $0<t\leq5$ Myr.}
\label{mig}
\end{figure}

Figure \ref{mig} illustrates how planets migrate in a specific MDW disk. For $r>1.2$ au, where $\Sigma(r)$ decreases
with $r$ (Fig. \ref{prof}), planets migrate inward. Below, for $r<1.2$ au, planets with $M_{\rm p} \sim 0.1$-1 $M_{\rm Earth}$ 
can be pushed out by the unsaturated corotation torque. The corotation 
torque saturates for $M_{\rm p} > 1$ $M_{\rm Earth}$ and these massive planets would therefore migrate inward due to the 
dominant Lindblad torques. Ogihara et al. (2024) already argued that this setup would provide explanation 
for the migration of super-Earths and mini-Neptunes to close-in orbits. In addition, the migration pattern
shown in Fig. \ref{mig} could help to concentrate terrestrial protoplanets near the zero-torque radius
and help with the RMC problem (Section 4.5).

For example, if protoplanets form and rapidly grow at $r<1$ au, they eventually reach the regime of outward migration
and are carried to $\sim 1$ au (for $M_{\rm p} = 0.1$-1 $M_{\rm Earth}$). As the migration of the outer protoplanets
slows near the zero-torque radius, the inner planets can then catch up. The behavior may lead to chains of protoplanets
in orbital resonances. The chains can survive to the end of the gas disk stage or be broken by stochastic perturbations
from gas turbulence (e.g., Rein 2012). A consequence of the latter case can be mergers between pairs of protoplanets
during the disk stage.

Additional complications arise from the time dependence of $\Sigma_{\rm g}$. The hydrodynamical simulation of
Suzuki et al. (2016) showed that early MDW disks are relatively flat, and the maximum of $\Sigma_{\rm g}$ tends to 
shift to larger distances for aging, low-density disks. We ignored such possible dependencies in the base 
models tested here (i.e., the $\beta_1$ and $\beta_2$ parameters are unchanging with time; Section 4).
If the surface density maximum shifted to larger orbital radii over time, the outer boundary
of the outward migration in Fig.~\ref{mig} would have moved to larger radii as well. At the same time, as the surface
density of gas disk decreased over time, Type-I migration slowed down. This suggests that the orbital architecture
of the terrestrial protoplanets in the gas disk could have been established relatively early, when the migration
torques were still relatively strong. 


\subsection{Stage 2: outer planet instability (5-15 Myr)}

For Stage 2, we took advantage of our previous simulations treating giant planet migration and instability
(e.g., Nesvorn\'y \& Morbidelli 2012, hereafter NM12; Deienno et al. 2017) to select a case that best satisfied
various Solar System constraints (see below). Jupiter and Saturn were placed on low eccentricity and low
inclination orbits in the 3:2 resonance at the beginning of Stage 1, and they remained in the 3:2 resonance
during Stage 1. We recorded the orbits of Jupiter and Saturn, as well as all terrestrial planetesimals/protoplanets
at the end of Stage 1 ($t=5$~Myr) to use as the initial conditions for Stage 2. Three ice giants were added
at $t=5$~Myr. They began in the resonant chain (3:2, 2:1, 3:2), a configuration that was previously shown
to work best to generate plausible dynamical histories (NM12; Deienno et al. 2017). The additional ice giant
had mass comparable to that of Uranus/Neptune.

\begin{figure}
\epsscale{0.8}
\plotone{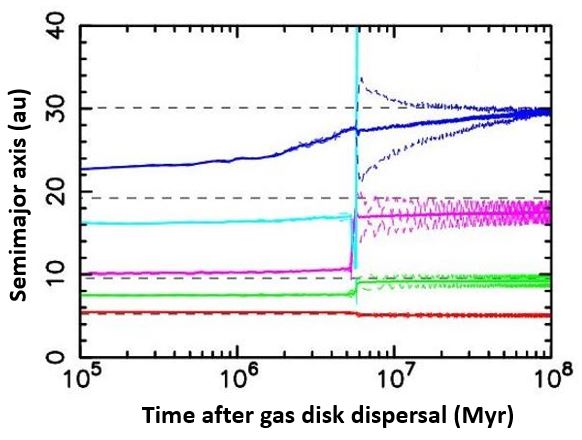}
\caption{The orbital histories of outer planets in our instability model. The planets started in the (3:2, 3:2, 2:1, 3:2) resonant 
  chain. The outer disk of planetesimals, not shown here, with the total mass $M_{\rm disk}=20$ $M_{\rm Earth}$ was placed beyond Neptune.
  The plot shows the semimajor axes (solid lines), and perihelion and aphelion distances (dashed lines) of each planet’s orbit.
  The black dashed lines show the semimajor axes of planets in the present Solar System. The third ice giant was ejected from the 
  Solar System during the instability about 5.8 Myr after the start of the simulation, which is 10.8 Myr after $t_0$
  in the timeline discussed in Sections 2.2 and 2.3.}
\label{case1}
\end{figure}

The properties of our dynamical instability model are illustrated in Fig. \ref{case1}. This model is identical to
Case 1 used in Nesvorn\'y et al. (2013), Deienno et al. (2014), and Roig \& Nesvorn\'y (2015). It satisfies
many Solar System constraints, with examples including the present orbits of the giant planets themselves,
the number and orbital distribution of Jupiter Trojans and irregular moons of the giant planets, and the
dynamical structure of the asteroid and Kuiper belts (see Nesvorn\'y 2018 for a review). Roig et al. (2016)
demonstrated, assuming that the terrestrial planets were already in place when the giant planet instability
happened, that the Case 1 model could explain the excited orbit of Mercury and the Angular Momentum Deficit
(AMD; Laskar \& Petit 2017) of our terrestrial planet system.

In the five-planet model,\footnote{The early Solar System is assumed to have five giant planets: Jupiter,
  Saturn, and three ice giants. NM12 showed that various constraints, such as the final orbits of outer
  planets, can most easily be satisfied if the Solar System started with five giant planets, with one ice giant
  ejected during the giant planet instability (Nesvorn\'y 2011, Batygin et al. 2012, Deienno et al. 2017).
  The case with four initial giant planets requires a massive planetesimal disk to avoid losing a planet, but
  that massive disk also tends to produce strong dynamical damping and long-range residual migration of Jupiter
  and Saturn that frequently violates constraints.} 
Jupiter and Saturn undergo a series of planetary encounters with the ejected ice giant. As a result
of these encounters, the semimajor axes of Jupiter and Saturn evolve in discrete steps. While the semimajor axis can decrease
or increase during one encounter, depending on the encounter geometry, the general trend is such that Jupiter moves inward,
i.e., to shorter orbital periods (by scattering the ice giant outward), and Saturn moves outward, i.e., to longer orbital
periods (by scattering the ice giant inward).

This process leads to the dynamical evolution known as the jumping-Jupiter model (Morbidelli et al. 2010). In the jumping-Jupiter
model, the principal coupling between inner and outer Solar System bodies occurs via secular resonances (e.g., $g_1=g_5$ and
$g_4=g_5$, where $g_j$ fundamental precession frequencies of planetary orbits; Brasser et al. 2009).
They cause strong orbital excitation, which in turn can inhibit accretional growth at specific radial distances
from the Sun (Clement et al. 2018, 2021a).

Cases with different kinds of giant planet instabilities were investigated in Clement et al. (2018, 2021a), Nesvorn\'y et al. (2021)
and Woo et al. (2024). Their results showed a general preference for an early instability (also see Agnor \& Lin 2012).
Accordingly, we assume here that the giant planet instability happened relatively early after gas disk dispersal; approximately
at $t=11$~Myr (Fig. \ref{case1}). This instability case was immediately available to us. Investigations of the two-source
model (Sections 4.8-4.10) with different instabilities is left for future work.  

Cases with delayed instabilities were studied in Woo et al. (2024). They found that if the last giant impact occurred at $t>80$~Myr,
the late accreted mass, defined as the mass accreted by the Earth after the Moon-forming impact (Morbidelli \& Wood 2015;
Section 3.2 here) is usually an order of magnitude lower than the value inferred from geochemical constraints. They therefore
suggested that the Moon formed at $50<t<80$ Myr. Using the noble gas argument from Nesvorn\'y et al. (2023), it can be inferred
that the delay between the giant planet instability and the giant impact that formed the Moon, $\Delta t$,
was $20<\Delta t<60$ Myr.
This interval would place the instability at $t<60$~Myr. Previously, the instability was constrained to $t<100$~Myr from the
survival of the Patroclus-Menoetius binary (Nesvorn\'y et al. 2018).  

The planetary positions and velocities in the selected instability model were recorded with a 1 yr cadence in NM12.
Our modified code {\it iSymba} (Roig et al. 2021), where {\it i} stands for interpolation, then reads the planetary orbits from a file,
and interpolates them to any required time subsampling (generally 0.01-0.015 yr, which is the integration time step used here
for the terrestrial planets). The interpolation is done in Cartesian coordinates. First, the giant planets are forward-propagated
on the ideal Keplerian orbits starting from the positions and velocities recorded by {\it Symba}  at the beginning of each 1 yr
interval. Second, the {\it Symba} position and velocities at the end of each 1 yr interval are propagated backward (again on the
ideal Keplerian orbits; planetary perturbations switched off). Third, the code calculates the weighted mean of these two Keplerian
trajectories for each planet such that progressively more (less) weight is given to the backward (forward) trajectory as time
approaches the end of the 1 yr interval. We verified that this interpolation method produces -- thanks to the high-cadence
(1 yr) sampling of the original instability simulations -- insignificant errors.

\subsection{Stages 3 and 4: residual migration (15-50 Myr) \\and the late stage (50-300 Myr)}

The Stage 2 integrations were run $t=15$ Myr. Much longer integrations were difficult to achieve with {\it iSymba} because
the interpolation method had large requirements on the computer memory and disk storage. As the giant planets were orbitally
decoupled from each other by the end of Stage 2, however, the continued integrations for $t>15$ Myr (Stage 3) did not need
to deal with planetary encounters. For $t>15$ Myr, it sufficed to use the standard {\it Symba} code and only account for
the slow, residual migration of the outer planets due to their interaction with planetesimals.

We implemented the planetesimal-driven migration/damping using artificial forces and adjusted the magnitude of these
effects such that the four outer planets ended up as close to their current orbits as possible. Stage 3 simulations were 
run to $t=50$ Myr, at which point the outer planets are already on their current orbits. We then continued all simulations 
with standard {\it Symba} to $t=300$ Myr to bring the terrestrial planet formation to completion.

\subsection{Collisional fragmentation}

It is debated whether debris generation in giant impacts is important for the terrestrial planet formation. On the one
hand, the Moon formed from a circumplanetary debris disk generated by a large impact on proto-Earth 
(Canup et al. 2023), and Mercury's mantle may have been stripped by a hit-and-run collision with a large body (Asphaug
\& Reufer 2014). On the other hand, Deienno et al. (2019) found no change in their terrestrial planet formation results
when fragmentation and different energy dissipation schemes were used to deal with collisions (also see Woo et al. 2024). 
Given that the main goal of our work is to 
test the effect of initial conditions and gas disk on terrestrial planet formation, in the bulk of our simulations, all 
collisions of protoplanets and planetesimals were assumed to lead to inelastic mergers. Our strategy was to sample 
different possibilities with simulations assuming mergers, identify the best cases, and re-run these cases with a more 
realistic treatment of collisions to see if there are any important differences.

\begin{figure}
\epsscale{0.7}
\plotone{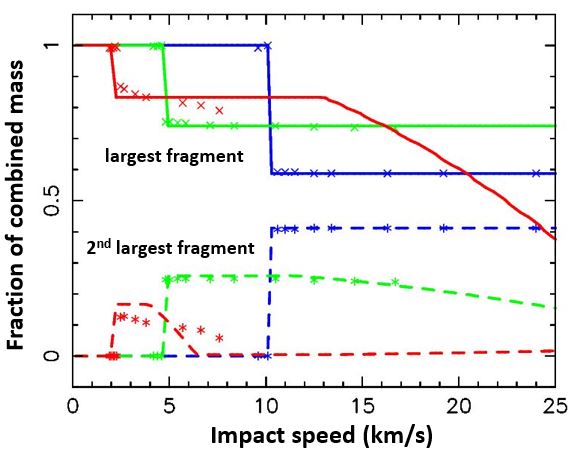}
\caption{A comparison between the scaling laws of Leinhardt \& Stewart (2012) and the SPH simulations 
of impacts reported in Emsenhuber et al. (2020). We considered three different collision setups for this illustration: 
the (1) target mass $m_{\rm target}=0.01$ $M_{\rm Earth}$, projectile mass $m_{\rm projectile}=0.2$ $m_{\rm target}$ and impact angle $\theta=45^\circ$
(red symbols and lines),
(2) $m_{\rm target}=0.1$ $M_{\rm Earth}$, $m_{\rm projectile}=0.35$ $m_{\rm target}$ and $\theta=60^\circ$ (green symbols and lines), and 
(3) $m_{\rm target}=1$ $M_{\rm Earth}$, $m_{\rm projectile}=0.7$ $m_{\rm target}$ and $\theta=75^\circ$ (blue symbols and lines). 
The solid lines show the mass of the largest fragment produced by collisions as a fraction of the combined 
mass, $m_{\rm largest}/(m_{\rm target}+m_{\rm projectile})$, as computed from Leinhardt \& Stewart (2012). This can be compared to 
the results of individual SPH simulations reported in Table 1 of Emsenhuber et al. (2020), labeled as crosses here.
The dashed lines and stars show the same for the second largest fragment. Abrupt changes of
$m_{\rm largest}/(m_{\rm target}+m_{\rm projectile})$ occur at the transitions between different fragmentation regimes.}
\label{lein}
\end{figure}

The fragmentation scaling laws were obtained from Leinhardt \& Stewart (2012). These laws account for many different collision
regimes, including the hit-and-run, super-catastro\-phic, erosive and partial-accretion cases. We programmed our own code and
verified that our results were identical to {\it Collresolve} from Cambioni et al. (2019) (with the collision model option
set to Leinhardt \& Stewart 2012). The results were compared to and found consistent with the SPH simulations 
of Emsenhuber et al. (2020, 2024). Figure \ref{lein} illustrates the results for different collisional regimes. 

The collisional code was included in {\it Symba} as follows. First, when a collision between two objects was identified,
we calculated the target and impactor masses as well as the impact speed and impact angle (defined as the angle $\theta$
between the relative impact speed vector at the time of impact and the normal vector to the surface element of first
contact). These parameters were used as an input for the collisional subroutine. 

Second, the collisional subroutine returns the mass of the first and second largest remnant bodies from the collision,
the mass of the smaller fragments, and a flag indicating the impact regime. We could not
afford, due to CPU limitations, to create many additional bodies in the $N$-body simulations.  As a compromise, we therefore
kept the first and second largest bodies in the simulation and removed the mass corresponding to the small fragments
(defined as $<10^{-4}$ $M_{\rm Earth}$). This procedure exaggerates the mass eliminated from the system, given that small 
fragments could potentially accrete on protoplanets at some later time (if kept in the simulation). In this sense, our 
simulations only approximate the potentially complex effects of collision fragmentation. 

Depending on the nature of the collision, different prescriptions were used for the post-impact velocity vectors of
the largest and second-largest bodies. 
For hit-and-run impacts, we assumed that the second-largest body continues its path along the
original velocity vector of the impactor. The velocity vector of the first largest body was then computed from the linear 
momentum conservation. For disruptive impacts, the second largest body was ejected from the surface of the first largest 
body at a velocity that slightly exceeded the combined escape speed. The velocity vector was directed out in the normal
direction to the surface element of first contact. The largest body bounced back due to the linear momentum 
conservation.\footnote{Special provisions were implemented in {\it Symba} such that the new bodies stay in their 
recursive level (Duncan et al. 1998) and freely float on divergent trajectories back to space (e.g., 
the past-step collision detection in {\tt symba7\_merge()} must be temporarily switched off).}     

\subsection{Additional effects}

The stochastic turbulent forcing of the gas disk on planets was included in some simulations. We added this effect
because we noticed that some level of stochastic forcing slightly helps to improve the success rate in some models
(see Sections 4.8 and 5 for an explanation). This affect is not fundamental, however, for the success of different
models and we did not investigated it as much as the influence of initial conditions, migration parameters, etc.
A more detailed investigation of the effects of stochastic forcing is deferred to future studies. 

The turbulent forcing was implemented following the algorithm described in Rein \& Choksi (2022) as a correlated
noise with a user-specified amplitude and auto-correlation time. Simulations of magneto-hydrodynamic turbulence found
autocorrelation times that were comparable to the orbital period (Oishi et al. 2007, Rein \& Papaloizou 2009). The
standard deviation of the force $\kappa$ was set relative to the gravitational force from the Sun. For example,
$\kappa \sim 10^{-6}$ implies that the typical magnitude of the stochastic force was a million times weaker than Sun's
gravitational force. The auto-correlation function was modeled as an exponential with an e-folding timescale $\tau_\kappa$.
We set $\tau_\kappa$ equal to the orbital period. A detailed description of stochastic force implementation in
Cartesian coordinates is given in Rein \& Choksi (2022). Note that our integration timestep is always much
shorter than the autocorrelation time, as required.


               
\section{Constraints}

A good terrestrial planet formation model includes enough physical effects to match constraints, and, if possible,
to make predictions. We discussed our model's physical effects in Section 2. Our task now is to identify the most
important and relevant constraints for the problem at hand. In the following sections, we will consider both
dynamical (Section 3.1) and cosmochemical/geophysical constraints (Section 3.2). 


\subsection{Dynamical constraints}

Different dynamical criteria were used to evaluate the success of our simulations. To quantify the radial distribution 
of planetary mass, Chambers (2001) defined the radial mass concentration, or RMC, as
\begin{equation}
S_{\rm c}={\rm max}\left( { \sum_j m_j \over \sum_j m_j[\log_{10}(a/a_j)]^2 } \right)\ ,
\end{equation}
where $m_j$ and $a_j$ are the mass and semimajor axis of planet $j$, the sum goes over all planets in the
inner Solar System (the outer planets were excluded), and the maximum is taken over $a$. Larger values of $S_{\rm c}$ 
indicate that the mass is more tightly packed around $a$. In the real Solar System, $S_{\rm c}=89.9$.

The angular momentum deficit or AMD (Laskar \& Petit 2017) was used to quantify the excitation 
of planetary orbits in eccentricity and inclination. The AMD is defined as
\begin{equation}
S_{\rm d}= { \sum_j m_j \sqrt{a_j}(1-\sqrt{1-e_j^2}\cos i_j) \over \sum_j m_j \sqrt{a_j} } \ ,
\end{equation}
where $e_j$ and $i_j$ are planetary eccentricities and inclinations. It measures the specific angular momentum difference
between perfectly circular and coplanar orbits, and that of the model orbits.  $S_{\rm d}=0.0018$ for the real
terrestrial planets. 

The advantage of these parameters is their objective mathematical formulation and relationship to the dynamical stability 
of planetary systems (e.g., Petit et al. 2018). Note, however, that they can be computed, and are often reported as
such, for simulations that do not produce meaningful results (when the number and/or masses of the final terrestrial
planets are incorrect). This can bias the statistical interpretation of results because correlations between
different measures exist (e.g., Deienno et al. 2019). For these reasons, we report the $S_{\rm c}$ and $S_{\rm d}$ statistics
only for the simulations that produced {\it good} planets.

To this end, we collected all planets with mass $m>0.5$ $M_{\rm Earth}$ between 0.5 and 1.2 au and called them Venus/Earth analogs.
The successful simulations were required to have exactly two Venus/Earth analogs (denoted by indices 1 and 2 below),
and no additional planets with $m<0.5$ $M_{\rm Earth}$ and $0.5<a<1.2$ au.

We also computed the radial separation between good Venus/Earth planets as 
$\Delta a=a_2-a_1$ (this is a complementary parameter to $S_{\rm c}$). In the real Solar System, Venus and Earth have 
$\Delta a=0.277$ au. Additionally, we determined, as a complement to $S_{\rm d}$, the mean eccentricities/inclinations of 
Venus/Earth analogs: $\langle e,i \rangle =(e_1+i_1+e_2+i_2)/4$. The real Earth and Venus have $\langle e,i \rangle =0.0274$.

The parameters $\Delta a$ and $\langle e,i \rangle$ measure the radial separation and orbital excitation of the two
large planets that form at 0.5-1.2 au. The $S_{\rm c}$ and $S_{\rm d}$ parameters are related to that but, unlike $\Delta a$ 
and $\langle e,i \rangle$, they are influenced by whether Mars and/or Mercury form in the simulations, and if so,
whether they at least approximately have the right mass (Clement et al. 2023).
For example, massive Mars/Mercury on perfect orbits would increase AMD/decrease RMC not because the orbits are too
excited/spread but because the planets are too massive. 

In this work, the planets with $a<0.5$ au were called Mercury and the planets with $1.2<a<1.8$ au were called Mars. We 
defined good Mercury analogs as planets with the mass $0.025<m<0.2$ $M_{\rm Earth}$ (i.e., mass roughly between a half and 
four times that of real Mercury, $M_{\rm Mercury}=0.055$ $M_{\rm Earth}$), 
and good Mars analogs as planets with the mass $0.05<m<0.2$ $M_{\rm Earth}$ (i.e., roughly between a half and a 
double of the actual Mars mass, $M_{\rm Mars}=0.107$ 
$M_{\rm Earth}$). We allowed for a larger mass range in the case of Mercury to leave space for the
possibility of mantle removal in a hit-and-run collision (Chambers 2013, Asphaug \& Reufer 2014, Clement et al. 2019;
recall that the bulk of our simulations adopted perfect mergers).

We defined bad Mercury and bad Mars analogs as planets with the mass $m>0.2$ $M_{\rm Earth}$. Bad Venus/Earth analogs were
assumed to be planets with $m<0.5$ $M_{\rm Earth}$. 

The good planetary systems identified in this work had {\it exactly two good Venus/Earth planets, one good Mercury, one
  good Mars, and no bad planets}. This is a very strict criterion. If, for example, each of the three criteria (Venus/Earth,
Mercury and Mars), is satisfied in 50\% of cases, and the results are not correlated, we would expect model success
in 12.5\% of all trial runs. This is the target probability we are aiming for in our analysis below.  

To separate the issue of Mercury and Mars formation, we will often discuss them below in the following context. Good
systems with Mercury alone, thereby ignoring any Mars criteria, and good systems with Mars alone, ignoring any Mercury
criteria. In this circumstance, our target probability would be 25\%. Note that no main asteroid belt
constraints were considered in this work.

\subsubsection{Summary} 

In Section 4, we will grade our simulations against the orbits and masses of the terrestrial planets. As described above,
we define good matches by four primary constraints:
\begin{enumerate}
\item Good Venus/Earth. Exactly two planets with semimajor axes $0.5<a<1.2$~au and masses $m>0.5$ $M_{\rm Earth}$, as well
  as no additional planets with $0.5<a<1.2$ au and $0.05 < m < 0.5$ $M_{\rm Earth}$ (i.e., no small Venus or small Earth). 
\item Good Mars. A single planet with $1.2<a<1.8$ au and $0.05<m<0.2$ $M_{\rm Earth}$. A common example of a bad Mars in
  our runs would have $1.2<a<1.8$ au and $m>0.2$ $M_{\rm Earth}$.
\item Good Mercury. Exactly one planet with $0.3<a<0.5$ au and $0.025 < m < 0.2$ $M_{\rm Earth}$, as well as no planet with
  $0.3<a<0.5$ au and $m>0.2$ $M_{\rm Earth}$. In other words, a very large Mercury would fail this criterion.\footnote{We
  also required that there were no protoplanets with $a<0.3$ au at the end of our simulations but this criterion was
  easily satisfied in all cases discussed in this work.}
\item Good Venus/Earth separation. It should have $\Delta a < 0.3$ au, with $\Delta a=0.277$ au for the real planets.
\end{enumerate}
In addition, for simulations that satisfied these criteria, we also considered additional constraints such as the AMD,
accretion history of planets, properties of the last giant impact, and various cosmochemical and geophysical measurements.

\subsection{Cosmochemical and geophysical constraints}

\subsubsection{Earth and Moon}

Many refractory elements have had their isotopic anomalies measured in meteorites and lunar rocks. Unlike most meteorites,
however, lunar rocks have isotopic compositions that are essentially indistinguishable from terrestrial composition
within current uncertainties. 
For example, the Earth and
Moon have $\Delta^{17}$O values within $\sim 10$ ppm of each other. In contrast, meteorites span a 
range of $\Delta^{17}$O values from $-4700$ ppm (Eagle Station Pallasite) to $+2600$ ppm (R chondrites), 
a factor of several hundred times larger range than the Earth-Moon difference (Dauphas \& Schauble 2016, 
Dauphas 2017). The Moon is therefore distinctively Earth-like in its isotopic anomaly composition.

One possible explanation for such similarity is that the impactor responsible for the Moon’s formation,
often called Theia, was made in the same isotopic reservoir of that of the Earth. It has also been argued
that the Earth-Moon isotopic similarity could reflect isotopic equilibration between the terrestrial magma
ocean and protolunar disk via the vapor phase (Pahlevan \& Stevenson 2007). A third option is that Theia
was large enough to produce extensive mixing between the proto-Earth and the protolunar disk (Canup 2012).
The limiting case of collision by two objects, each having one-half the Earth's mass, would produce
a disk and final planet that both contain 50\% target and 50\% impactor material. The proto-lunar disk
and planet would thus have equal compositions.  Finally, a fourth option is that the pre-collision
proto-Earth was spinning so fast that a relatively small Theia impactor could yield a protolunar
disk made predominately from the proto-Earth's mantle (Stewart et al. 2012).   

There is a considerable uncertainty in the timing of Moon formation, with a plausible range of
$\sim 30$-150 Myr (Fischer \& Nimmo 2018, Jacobson et al. 2014, Thiemens et al. 2019, Gaffney et al.
2023, Nimmo et al. 2024). Existing measurements have been interpreted to imply that the Earth and
Moon had equal W isotopic compositions when the Moon formed. The small 182W excesses in the lunar mantle
would then be explained by disproportionate late accretion onto the Moon and Earth (Bottke et al.
2010, Day \& Walker 2015). This scenario is supported by the evidence for a `young Moon' from Sm-Nd model
ages, which indicate lunar crust formation ages that vary from 4.35 to 4.45 Ga (Borg et al. 2015, 2019).
The time of lunar magma ocean crystallization is uncertain, with some evidence supporting melting and
global differentiation $\sim 200$ Myr after $t_0$ (Nimmo et al. 2024).

Highly siderophile elements (HSEs) such as Re, Os, Ir, Ru, Pt, Rh, Pd and Au, are strongly depleted
in the Bulk Silicate Earth (BSE; Chou 1978, Chou et al. 1983) due to their partitioning into the core
during the Earth differentiation epoch. Nevertheless, HSEs
are found in an approximately chondritic ratio in the primitive upper mantle (Chou 1978).
The most plausible explanation for this is the addition of a small amount of material of
approximately chondritic composition to the upper mantle after core segregation had ceased
(Chou 1978, Kimura et al. 1974, Dauphas \& Marty 2002, Rubie et al. 2016). It is estimated that this HSE-inferred late accretion (sometimes called
the Late Veneer) represented $\sim 0.5$\% of Earth mass that was accreted by the Earth and did not
re-equilibrate with the core metal (Turekian \& Clark 1969).

Morbidelli \& Wood (2015) discussed how the HSE-inferred late accretion constrains the amount of material
accreted by the Earth {\it after} the Moon-forming event. They argued that the HSE-inferred late accretion
probably means that the Earth accreted 0.25-1\% of its mass after the Moon-forming event. This relatively
small amount of accreted material was previously suggested as evidence that the Moon formed relatively late
(Jacobson et al. 2014), but Woo et al. (2024) revised this estimate to argue for the Moon formation
at 40-80 Myr after $t_0$.  

Finally, important constraints on Earth's accretion history can be obtained from the chemical
composition of its mantle. Previous work modeled the chemical evolution of the Earth's mantle
during a series of metal-silicate partial equilibration events associated with accretional
collisions (Rubie et al. 2011, Dale et al. 2025). The results imply that it is difficult to
simultaneously match the SiO$_2$ and FeO concentration in the BSE if the Earth accreted from
a uniformly reduced or a uniformly oxidized reservoir, or some mixture of the two.

The solution
to this conundrum favored in these studies was that the Earth initially accreted $\sim 70$\% of
mass from a reduced reservoir and subsequently accreted $\sim 30$\% of material from an oxidized reservoir
(probably similar to Ordinary Chondrites, OCs, or non-carbonaceous iron meteorite parent bodies;
Grewal et al. 2025). The reduced reservoir would have to have oxidation similar to that of ECs but different chemical composition 
(much larger Al/Si and Mg/Si ratios; Morbidelli et al. 2020) and different isotopic ratios (Burkardt et al.
2021) -- unsampled in the current meteorite collection. This inference would be difficult to interpret in the
dynamical models where the Earth accretes from a {\it single} reservoir (annulus or ring; Hansen 2009,
Woo et al. 2023, 2024, Dale et al. 2025). This constraint supplies some justification for the two-source
model discussed in Sections 4.8 and 4.9.  

\subsubsection{Venus}

Jacobson et al. (2017) considered constraints from the absence of internally-generated
magnetic field on Venus. They pointed out that whether or not a terrestrial planet can
sustain an internally generated magnetic field is determined by its initial thermodynamic
state and compositional structure, both of which are set by the early processes of accretion and 
differentiation. They showed that the cores of Earth- and Venus-like planets would grow with stable
compositional stratification unless disturbed by late energetic impacts. If a late energetic impact
occurred, however, it could potentially stir the core enough to create a long-lasting geodynamo. 
Jacobson et al. (2017) hypothesized that the accretion of Venus was characterized by the absence of
late giant impacts and the preservation of its primordial stratification. The absence of late giant impacts
on Venus would also be consistent with the fact that Venus does not have a moon (Jacobson \& Dobson 2022),
but note that Burns et al. (1973) already argued that a massive moon of Venus -- if it formed --
would be removed by tides.  

\subsubsection{Mars}

The accretion history of Mars can be elucidated from isotopic anomalies (Dauphas et al. 2024 and the 
references therein). Experimental analyses indicate that Mars' isotopic composition is consistently 
different from the Earth in terms of nucleosynthetic anomalies in Ti, Cr, Fe, Zn, Mo and O (Mezger 
et al. 2020). Among the elements displaying isotopic anomalies, siderophile elements are particularly 
valuable for constraining the late stages of accretion. Dauphas et al. (2024) measured the Fe isotopic 
compositions of several Martian meteorites, providing a more precise definition of the isotopic nature 
of Mars-accreted material. Their results suggest that Mars is a isotopic mixture of $\sim 65$\% EC
material and $\sim 33$\% OC material.\footnote{Liebske et al. (2025), considering a wide range of
  isotopic, geochemical and geophysical properties of Mars, found that Mars is unlikely to have formed
  from known unmodified meteoritic material. They suggested that the relatively oxidized building blocks
  underwent evaporation/condensation processes that
  lead to volatile-element depletion patterns unlike those in any known meteorite group.}
In contrast, the Earth probably accreted $\sim 95$\% of EC material (Dauphas et al. 2024).\footnote{Dauphas
  et al. (2024) assumed that there were no {\it unsampled} reservoirs. If one postulates the existence
  of an unsampled reservoir, the Earth composition would possibly be combined with a relatively large
  contribution of the unsampled reservoir, which would likely have many of the properties of the EC
  reservoir. This would be more in line with the chemical constraints from Rubie et al. (2011) and
  Dale et al. (2025).}

Dauphas et al. (2024) inferred that Mars began accreting a mix of OC and EC, but predominantly accreted
EC later than OC. It would seem natural to assume that the protoplanetary disk was radially stratified,
with EC material residing closer to the Sun than OC materials. The boundary between EC and OC may have been
located beyond Mars' present orbital radius, at $r>1.6$ au. Dauphas et al. (2024) suggested that Mars formed 
relatively early in a protoplanetary gas disk (see below) beyond 1.6 au. From there, it migrated inward
by gas-driven torques (Section 2.2.1) and eventually crossed the OC/EC boundary. This behavior
could explain the measured isotopic trend. 

182Hf-182W systematics of SNC (Shergotty-Nakhla-Chassigny) meteorites constrain the time of Mars’ formation.
Specifically, the measurements suggest that Mars reached approximately half of its present size within
$1.8^{+0.9}_{-1.0}$ Myr after CAIs (Dauphas \& Pourmand 2011, Kobayashi \& Dauphas 2013). These results
are consistent with the 60Fe-60Ni systematics of SNCs, which suggest that Mars reached 44\% of its present
mass in less than $1.9_{-0.8}^{+1.7}$ Myr after CAIs (Tang \& Dauphas 2014). These results imply that Mars
formed relatively rapidly when the protoplanetary gas disk was still around. Accounting for heterogeneities
in the Mars mantle produced by very large impacts would allow for somewhat longer accretion timescales
(Marchi et al. 2020). 

Additional evidence for the fast accretion of Mars comes from the isotopic signatures of the Martian atmosphere
that indicate the atmosphere came from nebular gas (P\'eron \& Mukhopadhyay 2022). The krypton and xenon isotopes
found in the Martian meteorite Chassigny, which are thought to represent Mars' interior, are inconsistent with the
atmosphere, suggesting that the atmosphere is not a product of magma ocean outgassing or fractionation of interior
volatiles. Atmospheric krypton instead originates from accretion of solar nebula gas after formation of the mantle
but before nebular dissipation. The implication is that Mars had to reach a large size at early enough times
that it could grab gas from the solar nebula, within $\sim 5$ Myr after $t_0$ (Section 2.2). 
 
\subsubsection{Mercury} 

Mercury has an anomalously large metallic core representing $\simeq 70$\% of Mercury's mass (Hauck et al. 2013).
Asphaug \& Reufer (2014) conducted hydrocode simulations of hit-and-run impacts on Mercury and showed that
proto-Mercury could have been stripped of its mantle in one or more high-speed collisions with a larger target
planet that survived intact. For this to work, Mercury would have to be one of many small protoplanets that
randomly avoided being accreted into larger bodies.

Clement et al. (2023) evaluated the core mass fractions (CMFs) of Mercury analogs obtained in their simulations of
terrestrial planet accretion with disruptive collisions. They favored a scenario where Mercury formed
through a series of violent erosive collisions between several, roughly Mercury-mass embryos in the inner
part of the terrestrial disk. These results are consistent with those of Franco et al. (2025), who used
numerical hydrocode simulations to show that collisions of similar-mass bodies can form a Mercury-like planet
if they have suitable impact angles and velocities. The simulations of Clement et al. (2023), however, failed
to produce the correct mass of Mercury. Moreover, Scora et al. (2024) adopted accretion conditions that would
be favorable for high CMFs of Mercury to form (excited orbits and high impact speeds), but obtained CMFs larger
than 0.4 in only $\sim 10$\% of their simulations with disruptive collisions. We address this issue in Section 4.10.

{ 

\subsection{Implementation of different constraints}

The four dynamical constraints described in Section 3.1.1 are considered as primary constraints in this work. These 
constraints are not subject to significant uncertainties and any reasonable model of the terrestrial planet formation 
should satisfy them. As different models tested here give different success probabilities in matching these criteria, 
the primary constraints are used for the model selection and parameter estimation. The excitation of planetary orbits, 
as defined by $S_{\rm c}$ or $\langle e,i \rangle$, is not considered to be a primary constraint because we observe that 
many different models lead to similar excitation of planetary orbits. We make sure that the orbital excitation is 
correct only for models that satisfy the primary criteria (e.g., Section 4.8).

The cosmochemical and geophysical constraints discussed above are not used for model selection. This is because many 
of these constraints have significant uncertainties (e.g., the timing of Moon-forming impact or late accretion of HSEs), 
are subject to different interpretations (e.g., the isotopic similarity of the Earth and Moon), and/or may be model 
dependent (e.g., the chemical composition of Earth's mantle). Once we select a preferred model, however, we consider 
all constraints discussed in Section 3.2. For example, the two-source model advocated in Section 4.8 implies isotopic 
similarity of the Earth and Theia (Section 3.2.1), is consistent with the chemical composition of Earth's mantle 
(Section 3.2.1) and isotopic difference between the Earth and Mars (Section 3.2.3). In Section 4.9, we show that the 
two-source model gives roughly the correct growth timescales for the Earth and Mars, plausible timing of the Moon-forming impact, 
and reasonable mass in the late accreted HSEs. The implications of collisional fragmentation for Mercury's mass and core 
(section 3.2.4) are discussed in section 4.10.   

}

\section{Results}

{We performed a large parametric study of different terrestrial planet formation models. Here we start by discussing 
the models that ignore effects of the protoplanetary gas disk (i.e., no Stage 1). The annulus model of Hansen 
(2009) is studied in Section 4.1. We show that this model, as proposed, has difficulties to form Mercury and 
produce a correct radial separation between Venus and Earth. In Sections 4.2 and 4.3, we therefore investigate -- 
still ignoring the gas disk effects --  whether different distributions of planetesimals at the end of the gas disk
stage could better satisfy these constraints. We find that a background planetesimal component extending down to 
$\sim 0.3$ au could help with the formation of Mercury (Section 4.3), but the problem with the Venus/Earth 
separation cannot be resolved in this setup.\footnote{All models discussed in Sections 4.1-4.3 do not account for
the giant planet instability and all models discussed in Sections 4.4-4.10 include the instability. We did not 
include the instability in Sections 4.1-4.3 because we wanted to closely reproduce the model of Hansen (2009) and investigate 
its slight modifications. The giant planets were placed on their current orbits at the beginning of these simulations.
We also repeated some of the simulations sets from Sections 4.1-4.3 {\it with} the instability and identified no 
significant differences in the results (also see Nesvorn\'y et al. 2021). The effects of the giant planet instability 
are much more important with the gas disk stage (Sections 4.4-4.10). where planets migrate into resonant chains and 
would often remain in those chains in the absence of an external trigger.}

Several models starting with the gas disk stage are discussed in Sections 4.4-4.10. We first test the planetesimal ring 
model with convergent migration (Section 4.4; Woo et al. 2024) and find that this model yields a relatively low probability 
to match our primary criteria (Section 3.1.1). Various subsequent modifications of this model show only a modest improvement 
(Section 4.5). We therefore investigate whether an {\it inner} planetesimal ring at $\sim 0.5$ au could help with Mercury's 
formation. We find encouraging results in this model, except for Mars that -- for obvious reasons -- is difficult to 
form from the inner ring. A follow-up model with the inner ring and planetesimal background extending to $>$1.5 au 
still yields only a 13\% success probability for Mars (Section 4.6).  
The radially extended planetesimal disks with convergent migration (Bro\v{z} et al. 2021) are ruled out in Section
4.7. 

Given these negative results, we therefore end up proposing a two-source model with the inner planetesimal ring at $\sim 0.5$ 
au and the outer source of planetesimals at 1.5-2 au (Section 4.8). We show that this model yields relatively high success 
probabilities for all primary criteria, and could match various cosmochemical constraints as well (Section 3.2). Implications
of the two-source model for planetary growth are discussed in Section 4.9. Additionally, we show that collisional fragmentation 
helps to improve chances to obtain a small Mercury from the inner ring (Section 4.10).
     
We are able to make these inferences mainly because we explored $\sim 100$ different models in total, and performed 
1000 simulations for each of them to fully understand the stochastic nature of the accretion process. This large parametric
study allows us carefully establish which of the explored models have better chances to match constraints. The
systematic sampling of model parameters is also the main reason why we are able to find better matches to the real 
terrestrial system than previous work. The preferred model with two source reservoirs is the best
one we were able to find -- all other models investigated here produce inferior results.}

\subsection{Annulus model}

The results of Hansen (2009) provide a useful reference case for the present study because their simulations
had a simple setup, small number of free parameters, and produced a relatively good match to the   
terrestrial planet system (as discussed below). That is why we consider this model here despite the fact that 
Hansen (2009) ignored the protoplanetary gas disk stage (which is nonphysical; also see Walsh \& Levison 2016).
Our models with the gas disk stage are discussed in Sections 4.4-4.10.

Hansen (2009) started their simulations with 400 (fully interacting) bodies randomly distributed in an annulus 
between 0.7 and 1 au. The total mass of protoplanets was set to 2 $M_{\rm Earth}$, which means that each initial 
body had the mass $\sim 0.4$ $M_{\rm Moon}$. All collisions were assumed to result in perfect mergers. Here we 
performed 1000 simulations of Hansen's model. Following Hansen (2009), for simplicity,
the gas disk effects were ignored.

In Hansen's setup, 652 of our 1000 simulations (65.2\%) produced good Venus/Earth analogs.\footnote{Recall that
  we define Venus/Earth analogs as planets with semimajor axes $0.5<a<1.2$~au and
  masses $m>0.5$ $M_{\rm Earth}$ (Section 3.1). A simulation produced good Venus/Earth analogs if the final 
terrestrial system contained exactly two planets that satisfied the above criterion and no additional planets 
with $0.5<a<1.2$ au and $0.05 < m < 0.5$ $M_{\rm Earth}$ (i.e., no small Venus or small Earth).}
This is an excellent success rate. As for Mars, of our 1000 simulations, 230 (23\%) simultaneously had a good
Venus/Earth, a good Mars ($1.2<a<1.8$ au and $0.05<m<0.2$ $M_{\rm Earth}$), and no examples of a bad Mars
($1.2<a<1.8$ au and $m>0.2$ $M_{\rm Earth}$). This means that -- for the systems with a good Venus/Earth
-- about $23/65.2=35$\% also had one good Mars and no bad Mars, a very good success as well. 

This cannot be the whole story, however, because in the Hansen model, Mars forms from the same annulus material
as the Earth and should presumably have the same isotopic composition as the Earth. In contrast, the
cosmochemical constraints discussed in Section 3.2.3 indicate that Mars and Earth have distinct isotopic
compositions (e.g., Mezger et al. 2020). This shows the need for Mars to accrete additional material from
another isotopic reservoir. We discuss different scenarios to explain this possibility below.

Unfortunately, Hansen's model does not work well for Mercury. Out of 1000 simulations, we found that only
56 (5.6\%) simultaneously had a good Venus/Earth and a good Mercury.\footnote{Recall that model systems
  with good Mercury  
were defined here as having exactly one planet with $0.3<a<0.5$ au and $0.025 < m < 0.2$ $M_{\rm Earth}$, 
and no planet with $0.3<a<0.5$ au and $m>0.2$ $M_{\rm Earth}$ (very large Mercuries would fail this 
criterion).} This means that only $56/65.2=8.6$\% of cases with a good Venus/Earth also had a good Mercury. 
While we cannot exclude that the formation of Mercury was a low probability event, this outcome motivated us 
to examine additional scenarios, first still without the Stage 1 (Sections 4.2-4.3) and then with the disk 
stage included (Sections 4.4-4.9).\footnote{On closer inspection, we found that the Mercury-Venus 
separation is often incorrect in Hansen's model. Only 5 out of 56 simulations (9\%) with good Mercury, 
Venus and Earth give the Mercury-Venus separation $>0.3$ au (the real separation is 0.336 au). In contrast, 
in the two source model discussed in Section 4.8 ({\tt model203}), 55 out of 157 simulations (35\%) with 
good Mercury, Venus and Earth give the Mercury-Venus separation $>0.3$ au, which is more reasonable.}  

We investigated the accretion histories of planets in individual simulations to understand why they often
produce a good Mars and a bad Mercury. We found that good Mars analogs accreted from the 0.7-1 au annulus
and were scattered to $a>1.2$ au, where their orbits were circularized by collisions and gravitational
interactions with other planetesimals/protoplanets. This often produced Mars with the correct mass and
orbit. The problem with Mercury is that planets with $a<0.5$ au rarely formed in the annulus model.
Apparently, it is difficult for a small protoplanet to be scattered all the way from $a>0.7$ au down to
$a<0.5$~au and then survive on a stable orbit. In this sense, in Hansen's model, it is difficult to achieve
the relatively large radial separation between Mercury and Venus.

A related problem was identified for the radial separation of Venus and Earth, namely that only 19\% of
jobs with a good Venus/Earth had $\Delta a < 0.3$ au. This percentage is slightly misleading because when
an individual simulation ended with $\Delta a < 0.3$~au, the model Venus/Earth were often less massive
than the real planets, their orbital radii were smaller, and/or there was a third very massive planet
at $a>1.2$ au (super-Mars). When we evaluated $\Delta a$ for cases with a good Venus/Earth and a good Mars,
to control the influence of mass spreading beyond 1.2 au, and renormalized the orbits of Venus/Earth
to have the mean semimajor axis of these planets equal to the real value (0.86 au), the fraction of
good simulations with $\Delta a < 0.3$ au dropped to 5.7\%. This means that only 1 in $\sim 18$ simulations
reproduced the tight radial separation of Venus and Earth.

The root of the problem here is the gravitational scattering of growing protoplanets. As they grow and scatter,
they move from the parent annulus to smaller and larger orbital radii. This behavior is difficult to avoid,
such that the two major planets typically end up with $\Delta a > 0.3$~au. 

We therefore face a contradiction in Hansen's model. To get a good Mercury, we need small protoplanets to
be more strongly scattered inward, but to get Venus and Earth with a good separation, the scattering of large
protoplanets must be reduced. In addition, Mars needs to accrete material from a planetesimal reservoir whose
isotopic composition is distinct from the one that made most of Earth.

\subsection{A ring model}

Given the results above, we tested whether the radial separation of Venus and Earth could be reduced when 
planetesimals start in a very narrow annulus. This kind of architecture is often called the ring model (Woo et al.
2023, 2024), which was motivated by the proposed early formation of planetesimals at the silicate sublimation
line (Morbidelli et al. 2022). With this in mind, in a slight departure from Hansen (2009), the initial
planetesimals were assumed to follow a Gaussian ring distribution centered at $r_1$ with a narrow width
$\sigma_1$ (Eq.~\ref{ring}). We first tested $r_1=0.85$ au -- an intermediate value between the orbital
radii of Venus and Earth -- and $\sigma_1=0.1$ au (Woo et al. 2023). In the spirit of Section 4.1, to proceed
step by step from Hansen's model toward physically more complete models (Section 4.4 onward), the gas disk
effects were still ignored in this section.

The ring simulations produced the following success rates for a good Venus/Earth, a good Mars (and no bad Mars),
and a good Mercury; (V/E, Mars, Merc) = (62\%, 37\%, 7.2\%) (Table 1). These results also mean that 23\% of
our simulations in total produced a good Venus/Earth/Mars, while only 4.4\% produced a good Venus/Earth/Mercury.
These percentages are similar to those obtained for the original setup of Hansen (2009).

With the ring at 0.85 au, Venus and Earth often formed at somewhat smaller orbital radii than the real planets.
This presumably happened because shorter accretion timescales at small orbital radii favored accretion of large
planets closer to the Sun. After a renormalization for this shift (see previous section), we found that only
5.3\% of systems with a good Venus/Earth/Mars had $\Delta a < 0.3$ au. This outcome is similar to the original
setup. 

With $r_1=0.85$ au and $\sigma_1=0.03$ au, we obtained (V/E, Mars, Merc) = (63\%, 38\%, 7.7\%) (Table 1) --
almost no change from $\sigma_1=0.1$ au.  We also found 10\% for the Venus/Earth separation, modestly better
than the original case but still not very large. Ideally, we would like this fraction to approach 50\%. To make
these results easier for the reader to follow, we will define it as (V/E, Mars, Merc, V/E Sep) = (63\%,
38\%, 7.7\%, 10\%) (Table 1). Note that the use of even narrower rings ($\sigma_1<0.03$ au) did not improve
the situation because planetesimals quickly spread by gravitational scattering (see the next section). 

We also tested cases with a larger number of initial planetesimals, $N=1000$ in total, to see if improving
the resolution could affect the results, and different ring locations, $r_1=1$ au and 1.2 au, in an attempt to
generate more accurate orbital radii of Venus and Earth. The case with 1000 initial planetesimals produced results
very similar to those quoted for the original case above: (V/E, Mars, Merc, V/E Sep) = (53\%, 37\%, 15\%, 4\%)
(Table 1).  These results indicate that resolution may not be responsible for the problems discussed above.
The case with $r_1=1$ au gives the right orbital radii for Venus and Earth (mean 0.87 au, very close to the
real 0.86 au; the ring at 0.85 au gives the mean 0.78 au). The percentages quoted for the four success criteria
used above are (V/E, Mars, Merc, V/E Sep) = (49\%, 32\%, 3.9\%, 4.4\%) (Table~1). The case with $r_1=1.2$ au
generates the orbital radii of Venus and Earth that are too large (the mean 0.94 au). 

In summary, the annulus and ring models give reasonable success for Venus, Earth and Mars but fail in 
two important aspects: the (1) good Mercury analogs do not form very often (only 4-15\% of simulations 
with good Venus/Earth have good Mercury), and (2) orbital separation of Venus and Earth is often too large 
(90-95\% of cases have $\Delta a > 0.3$ au).  In addition, the ring would have to be centered at 
$r_1 \simeq 0.85$-1~au for Venus/Earth to grow at the right orbital distance.\footnote{Recall that we are 
  neglecting the gas stage here. Models with the disk migration of protoplanets allow for a larger range
  of ring locations (Section 4.6).}
A very narrow ring with $\sigma_1 \lesssim 0.03$ au slightly helps reduce the Venus/Earth separation but does
not resolve the problem (only $\sim 10$\% success). 

\subsection{Adding background}

Next, with an eye to the isotopic composition of Mars (Section 3.2.3), and Mercury, we tested models where the initial
distribution of planetesimals consisted of two components: the Gaussian ring and a radially extended background
(Section 2.1). We used $r_{\rm min}=0.3$ au, $r_{\rm max}=1.8$ au, $\gamma=0$ in Eq. (\ref{back}), and varied the
background contribution, $0 \leq w_3 \leq 0.5$, to obtain optimal results. The total mass of planetesimals was
fixed at 2.1 $M_{\rm Earth}$. The gas disk effects were still ignored.

We found that this setup substantially increased the success rate for Mercury. The best results were obtained
for $w_3=0.2$, where we had 47\% of simulations with good Mercury (again, the percentage quoted here is the fraction
of simulations with a good Venus/Earth that also produced a good Mercury and no bad Mercury; the model labeled
{\tt back20} in Table~1).  This value is $\sim 6.5$ times better than the case without background (only 7.2\%
success for Mercury with $w_3=0$; the {\tt ring4} model in Table 1). 

\begin{figure}
\epsscale{0.8}
\plotone{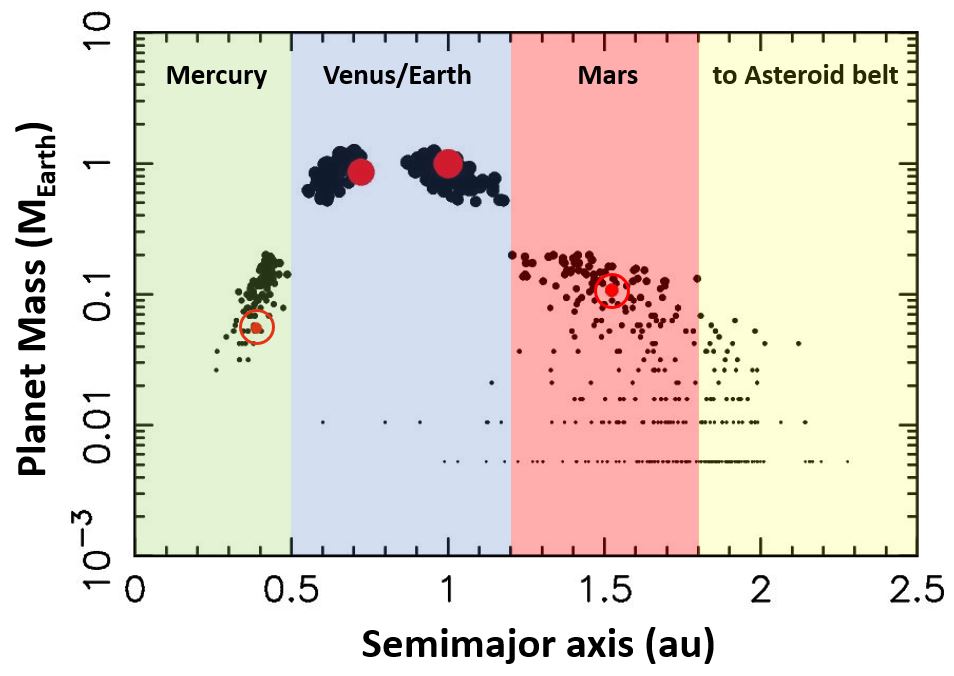}
\caption{The result of 100 successful simulations of the {\tt back20} model (the ring at 0.85 au with a 20\% background)
  that produced a good match to all terrestrial planets (as defined in Section 3.1). The black dots show the
  results of individual simulations with good final planets. The bigger red dots are the real planets. 
  Note that we do not exclude simulations that produced additional bodies with $m<0.05$ $M_{\rm Earth}$ in the Mars 
  region ($1.2<a<1.8$ au). Some of these bodies would collide with Mars at $t>200$ Myr (these simulations were terminated
  at 200 Myr).}
\label{h701}
\end{figure}

\begin{figure}
  \epsscale{0.7}
\plotone{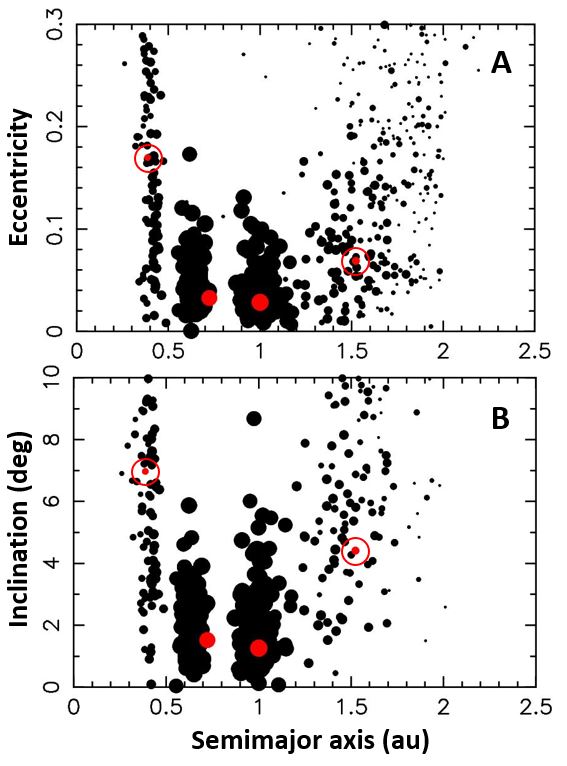}
  \caption{The result of 100 successful simulations of the {\tt back20} model (the ring at 0.85 au with a 20\% background)
  that produced a good match to all terrestrial planets (as defined in Section 3.1). The black dots show the
results of individual simulations with good final planets. The bigger red dots are the real planets.}
\label{orb1}
\end{figure}

The background is clearly beneficial for Mercury's formation (Fig. \ref{h701}). Overall, out of 1000 simulations,
100 (10\%) produced good planets as defined in Section 3.1. This success percentage is close to our target
of 12.5\%. The model orbits in the good simulations are compared to the real orbits in Fig. \ref{orb1}.

With a 20\% planetesimal background, typically $>50$\% of the material accreted by Mercury starts in the ring
at $r>0.7$ au. Collisions between the proto-Mercury and background objects help to stabilize the orbit of
proto-Mercury when it is scattered from the ring to $r<0.5$ au. The background below 0.5 au contributes
20-50\% to Mercury's total mass budget, with the simulations producing considerable variability.  With an
increased number of background objects, the fraction of simulations with a good Venus/Earth slightly decreases.
We do not consider this outcome to be particularly meaningful because the fraction of runs with a good Venus/Earth
remains relatively high ($\gtrsim 50$\%; Table 1). The presence of background objects also does not seem to
substantially affect the success rate for Mars (35-38\% in Table 1).\footnote{The results for the {\tt back20}
  model with the bulk density of all bodies set to $\rho=5$ g cm$^{-3}$ (Section 2.1) are very similar to those
  reported here for $\rho=3$ g cm$^{-3}$.}

\begin{figure}
\epsscale{0.7}
\plotone{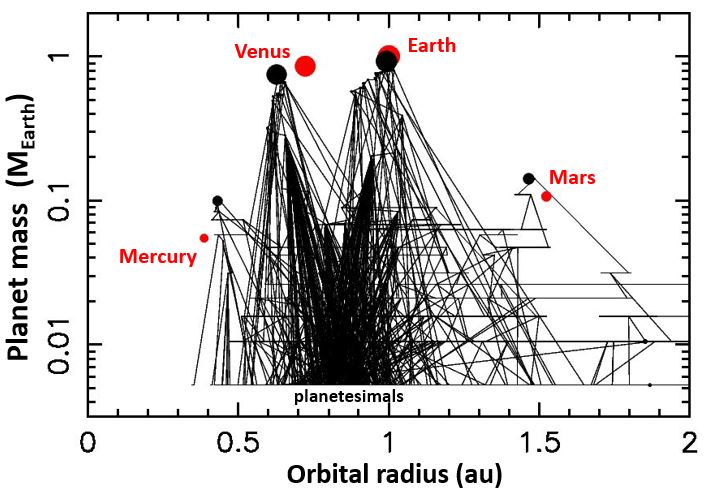}
\caption{The accretion history of planets in one of the best cases from the simulation set {\tt back20} (a ring at 
0.85 au, 20\% background, no gas). Each pair of connected line segments represents one accretion event. The black and red dots 
are the model and real planets, respectively. The dot size is proportional to planet mass.}
\label{christ}
\end{figure}

Motivated by the cosmochemical constraints discussed in Section 3.2, we evaluated the provenance of material
that contributed to each planet's mass budget. For that, in each run, we recorded the initial orbital radius
of planetesimals and followed the accretion sequence to determine their fate (Fig. \ref{christ}). This gave
us a full record of how planetesimals from different heliocentric distances were incorporated into planets.
For example, in the simulation set with $w_3=0.2$, we found that $\gtrsim 80$-95\% of Earth's material was
accreted from the ring below 1 au. This can be compared with the isotopic constraints discussed in Section
3.2.3, which indicate that the Earth accreted $\sim 92$\% of EC material in total (Dauphas et al. 2024;
the possibility of an unsampled reservoir is ignored here). In the context of the present single-ring
model, it would therefore be natural to associate the provenance of the EC material with the planetesimal
ring (Dale et al. 2025).

As for Mars, there is a large scatter among individual simulations indicating that $\sim5$-50\% of Mars’ mass
accreted from from $r>1$ au ($\sim 20$\% on average). This could have interesting implications for the radial
separation of the EC and OC reservoirs. Ignoring the possibility of an unsampled reservoir, the isotopic constraints
discussed in Section 3.2.3 would require $\sim 92$\% EC in Earth, $\sim 65$\% EC in Mars, and $\sim 32$\% OC
in Mars. If we assume that the EC/OC boundary was at the orbital radius $r_{\rm EC/OC}$, with EC material on inside and
OC material on outside, then $r_{\rm EC/OC} \sim 1$ au would best satisfy the constraints.\footnote{Spectroscopic
  studies classify many Hungaria asteroids at 1.8-2 au as E-type in asteroid taxonomy (Warner et al. 2009).
  These bodies are a spectral match to aubrites, meteorites whose high enstatite content, oxygen isotopic
  compositions, and formation under extremely reducing conditions suggest a kinship with enstatite chondrites
  (e.g., Keil 2010). If EC material formed at $r<1$ au, as the above arguments imply, Hungarias would have
  to be transported from $r<1$ au to $r>1.8$ (see, e.g., Bottke et al. 2006, 2012).}     
This value is roughly the orbital radius where our ring merges with the background population of planetesimals. 
These results are not ideal, however, as the average mass accreted by Mars from $r>1$ au is somewhat lower that 
implied by isotopic constraints.\footnote{Recall 
that the models discussed here and in Sections 4.1 and 4.2 do not include the effects of the giant planet 
instability. We repeated some of these simulations with the giant planet instability and did not identify any 
large differences in the radial mixing of materials accreted by individual planets.}

\begin{figure}
\epsscale{0.7}
\plotone{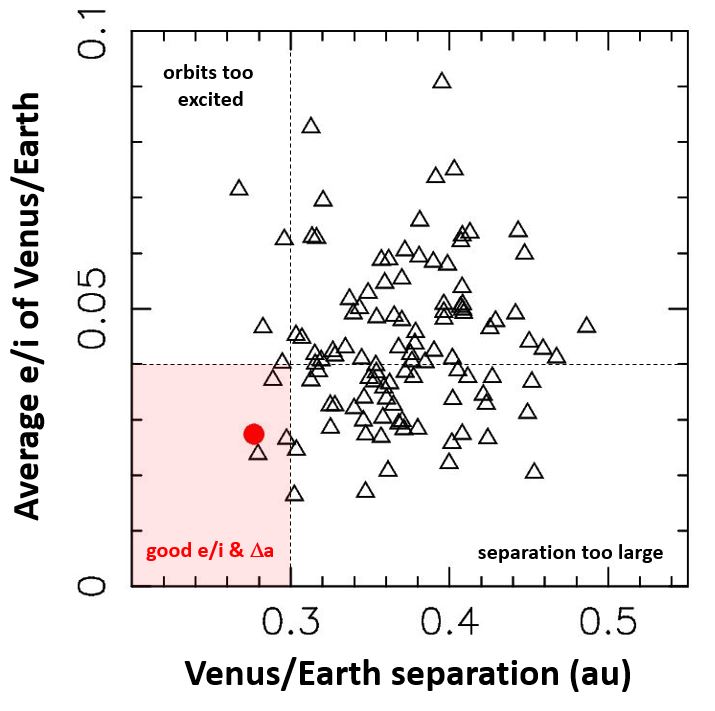}
\caption{The Venus/Earth separation, $\Delta a$, and the average eccentricities/inclinations,
  $\langle e,i \rangle$, for 111 successful simulations from {\tt 0.01au} (the narrow ring at 0.85 au with
  a 20\% background) that produced good terrestrial planets (as defined in Section 3.1). The triangles
  show the results of individual simulations with good final planets. The red dot is the
  real planets.}
\label{delta}
\end{figure}

Whereas some of these results are encouraging, we note that the problem with the radial mass concentration persists.
In our best case, with a very narrow ring ($\sigma_1=0.01$ au) and a 20\% background, only 7.2\% of successful
simulations with good planets had the Venus/Earth separation $\Delta a < 0.3$ au (Fig. \ref{delta}; the simulation
labeled {\tt 0.01au} in Table 1). It is not surprising that the addition of background objects did not help with
the separation issue, because it obviously did not provide any mechanism for bringing Venus and Earth closer
together. It may be the case that the separation between Venus and Earth is unusually tight, simply by chance,
for the accretion conditions in which these planets formed. Related to that, we found that the tight radial
separation of Venus and Earth was produced in some simulations
when a large projectile responsible for the last giant impact (Theia) had smaller orbital radius than the
Earth, and the Earth's orbital radius  moved sunward as a result of its impact. In this sense, the
nature of the last giant impact could be related to the RMC problem.\footnote{We also looked into simulations
  from Nesvorn\'y et al. (2021), specifically one of their most successful cases that started with 100
  lunar-mass protoplanets and a background disk of planetesimals. In this model,
  (V/E, Mars, Merc, V/E Sep) = (62\%, 17\%, 40\%, 9\%) (Table~1).  These percentages are similar to those
  obtained for the runs from this section for all but Mars, which is lower. With that said, the results
  from Nesvorn\'y et al. (2021) are subject to small number statistics because they only completed 100
  simulations for their best models.}    

\subsection{Previous models with gas disk effects}

The model of Hansen (2009), and its variations with the radially extended background, is not a complete
model of terrestrial planet formation. The reason is because planetesimals and protoplanets in the
terrestrial planet region formed and dynamically evolved in a protoplanetary gas disk (Scott \& Krot
2014, Budde et al. 2018, Spitzer et al. 2021, Piralla et al. 2023), a stage that was ignored in the
previous sections. To investigate these issues, we now turn our attention to models including the
protoplanetary disk stage (Section 2.2).

We first tested gas disks with surface densities that smoothly decreased with the orbital radius,
$\Sigma_{\rm g}(r)=\Sigma_0 (r/r_0)^{\beta}$ with $\Sigma_0=1700$ g cm$^{-2}$, $r_0=1$ au and $\beta=-0.5$, 
$-1$ and $-1.5$. For reference, the MMSN has $\beta=-1.5$ (Weidenschilling 1977, Hayashi 1981).
These tests were unsuccessful. For example, when the initial planetesimals were distributed in a ring
with $r_1=1$ au, $\sigma_1=0.1$ au, and $\tau=1$ Myr and $\beta=-1$, only 26\% of simulations produced
a good Venus/Earth, only 11\% a good Mars, and none a good Mercury.  In other words,
(V/E, Mars, Merc) = (26\%, 11\%, 0\%) (Table~1). This happened because relatively large protoplanets 
formed within the gas disk lifetime and migrated inward, often ending up at $r<0.5$ au. 

We do not discuss these cases further here because Woo et al. (2023) already documented this issue.
For things to work, planet growth within the gas disk would have to be slowed down to the point that
the protoplanets remained small ($\sim$ Mars-mass or smaller) and did not migrate.  They would probably
have to form slowly from an extended and/or low-mass planetesimal disk, perhaps because terrestrial planetesimals 
formed late during the gas disk lifetime (the results with the extended disks are not promising; Section 4.7). 
Alternatively, the gas disk would have to be relatively short lived ($\tau < 0.1$ Myr in the ring model 
with the fast growth of protoplanets). None of these possibilities, however, is supported by the existing
constraints (Section 3.2).

Woo et al. (2023, 2024) conducted simulations of terrestrial planet growth from a ring with $r_1=1$ au
and $\sigma_1=0.1$ au. They also ran simulations where the surface density of the gas disk was assumed to
peak near 1 au, inducing convergent migration of protoplanets toward 1 au. For Stage 1, they used the
GENGA code (Grimm \& Stadel 2014), which is an integrator similar to the one used here, but runs on GPUs. The simulations
were continued through Stages 2-4 using the setup identical to the one described in Sections 2.3
and 2.4. Their simulations had much a better resolution for the initial planetesimal disk, with
$N=8000$ planetesimals per simulation, but much smaller statistics as for the number of completed
simulations (typically 10 per case).

Here we conducted 300 simulations of their case {\it shallow} starting from one of their most favorable
results at the end of Stage 1. This set of simulations was analyzed with the same criteria as done
above. It yielded (V/E, Mars, Merc, V/E Sep) = (23\%, 9\%, 26\%, 0\%), which is unsatisfying.  

This outcome illustrates the difficulty in reproducing the terrestrial planet system with simulations
that include gas disk effects. Within the gas disk, planetesimal orbits become circularized by
aerodynamic gas drag before they can be scattered over large radial distances. Protoplanets
therefore have a diminished ability to scatter planetesimals away from the initial ring location
to smaller and larger orbital radii.

In addition, the orbits of protoplanets are strongly damped by disk torques, such that they cannot
be scattered very far either. This limits the ability of these simulations to produce Mercury and/or
Mars by scattering bodies away from the ring. Moreover, as the large protoplanets grow and migrate,
they accrete and severely reduce the population of smaller protoplanets and planetesimals, such
that the material available for scattering after the gas disk lifetime is too small to be
relevant.\footnote{The intermediate mass bodies with diameters $D \sim 1000$ km are the most
  susceptible to scattering during the gas disk stage because disk torques and gas drag are
  relatively weak for them.}
With these populations strongly reduced by the end of the gas stage, the late stage of accretion
between large protoplanets also tends to be more violent, which in turn yields a wider range of outcomes.

We emphasize that the growth of the terrestrial planets with and without the effects of protoplanetary
gas disk happens in fundamentally different dynamical regimes. The orbital eccentricities and
inclinations of bodies in a protoplanetary gas disk remain small, typically $\lesssim 0.01$. This
low excitation in turn creates a dynamical environment with fast initial growth, low impact velocities,
and the tendency of protoplanets to migrate into orbital resonances with one other. Without the gas
disk, the orbital eccentricities and inclinations become relatively large ($\sim 0.1$). This
increased excitation leads to a dynamical environment with lower impact probabilities, relatively
slow but sustained growth, and large impact velocities. 

\subsection{The ring model with convergent migration}

Our gas disks have a fixed radial structure and exponentially decay over time with $\tau=1$~Myr. To generate
convergent migration, the radial structure of gas disks is constructed following Eq. (\ref{sigma2}) with
$\Sigma_0=1700$ g cm$^{-2}$, $\beta_1=-1$ and $\beta_2=0$. This profile is more rounded than the `shallow' disk
in Woo et al. (2024) (Fig. \ref{prof}), but is likewise reminiscent of the MDW disks (Suzuki et al. 2016,
Kunimoto et al. 2020, Ogihara et al. 2024).
The migration map for this disk was shown in Fig. \ref{mig}. There is a broad region of masses and orbital
radii where protoplanets migrate outward. We start by discussing the results with $r_1=0.85$ au,
$\sigma_1=0.1$ au and $w_3=0.2$ (the background planetesimals contribute by 20\% to the total initial mass),
$r_{\rm min}=0.3$ au, $r_{\rm max}=1.8$ au and $\gamma=0$ in Eq. (\ref{back}) (model {\tt gas1} in Table 1).
This initial setup is identical to that used in Section 4.3, but now we also include the gas disk effects.

Using the usual success criteria (Sect. 3.1), we obtained (V/E, Mars, Merc, V/E Sep) = (39\%, 15\%, 36\%, 14\%).
This is a substantial improvement over the results from Woo et al. (2024), but the probabilities are significantly
lower than in our best model without gas (e.g., the 37\% success for Mars in {\tt back20}; Table 1). 

The success rate for Venus/Earth is lower here (39\%) than without gas (55\% in {\tt back701}). This difference
is probably a consequence of stronger stochastic effects during the late stage of planet accretion. During the
gas disk stage, several large protoplanets grow at $0.5<r<1.2$ au. These protoplanets accrete nearly all other
bodies at $r < 1.2$ au -- planetesimals and small protoplanets only survive in the Mars region at $r>1.2$ au.
Consequently, when the gas disk is gone and the large protoplanets start interacting with each other, there
are fewer small bodies to dynamically cool orbits. As for Venus/Earth separation, we found a slight improvement
over the results discussed above (14\% vs. $\lesssim10$\%), but these results are not terribly satisfying.

We investigated several modifications of the {\tt gas1} model.  They included disks with weaker/stronger
migration torques, turbulent disks, and various modifications of initial conditions (very narrow rings,
different background weights and profiles). None of these models represented a substantial improvement over
the case discussed above.\footnote{The giant planet instability is important in the terrestrial planet accretion
  models that include the gas disk stage. The instability helps to break resonant chains between protoplanets
  (established during their disk-driven migration) and trigger the late stage of giant impacts (Woo et
  al. 2024). This is less of an issue when the gas stage is ignored because the orbits of protoplanets 
  remain intrinsically unstable in this case (e.g., Nesvorn\'y et al. 2021).}    

\subsection{An inner ring for Mercury?} 

The planetesimal ring from which the terrestrial protoplanets grow does not need to be placed at 0.85-1 au
in a convergent disk with a density bump near 1 au. Instead, it can be placed at smaller
orbital radii, potentially even below 0.5 au. As protoplanets grow from the inner ring in our disk model,
they could migrate outward to $\sim 1$ au and concentrate there, potentially accreting into good Venus/Earth
analogs.\footnote{Note that that the planetesimal ring does not form at any strong pressure bump in Morbidelli
  et al. (2022). It forms at the silicate sublimation line which was probably located at $\sim 0.5$-1 au in
  the early protoplanetary disk (Marschall \& Morbidelli 2023). The planetesimal ring location therefore
  does not need to coincide with the maximum of the surface gas density.} 

The inner ring could also help to improve our chances of obtaining a good Mercury in the simulations. 
In the gas disk, where scattering over large radial 
distances is difficult and the growth of protoplanets is essentially local, massive protoplanets would accrete 
in the inner ring and migrate out. Mercury would not migrate that much because it is less massive (Fig. \ref{mig}). 
This raises the possibility that Mercury was one of the last small protoplanets forming at the inner ring -- after
more massive protoplanets already migrated out -- that was left behind near the original ring location.
The outward migration of massive protoplanets from the inner ring could produce a gap, both in mass and orbital
radius, between Mercury and Venus/Earth (Clement et al. 2021b).

The possibility of having an inner ring of planetesimals is in line with the existing models of temperature
profile in the protosolar nebula (Morbidelli et al. 2022, Izidoro et al. 2022). These models show that the
silicate sublimation line, where the first planetesimals presumably form, moves in the sunward direction
within an early disk. It can potentially arrive to $\sim 0.5$~au in $<0.5$ Myr (Morbidelli et al. 2022). In some
disk models with lower temperatures, the silicate sublimation line starts near 0.5 au (Marschall \& Morbidelli
2023). If so, it is plausible that a planetesimal ring formed near $r \sim 0.5$ au. 

\begin{figure}
\epsscale{0.8}
\plotone{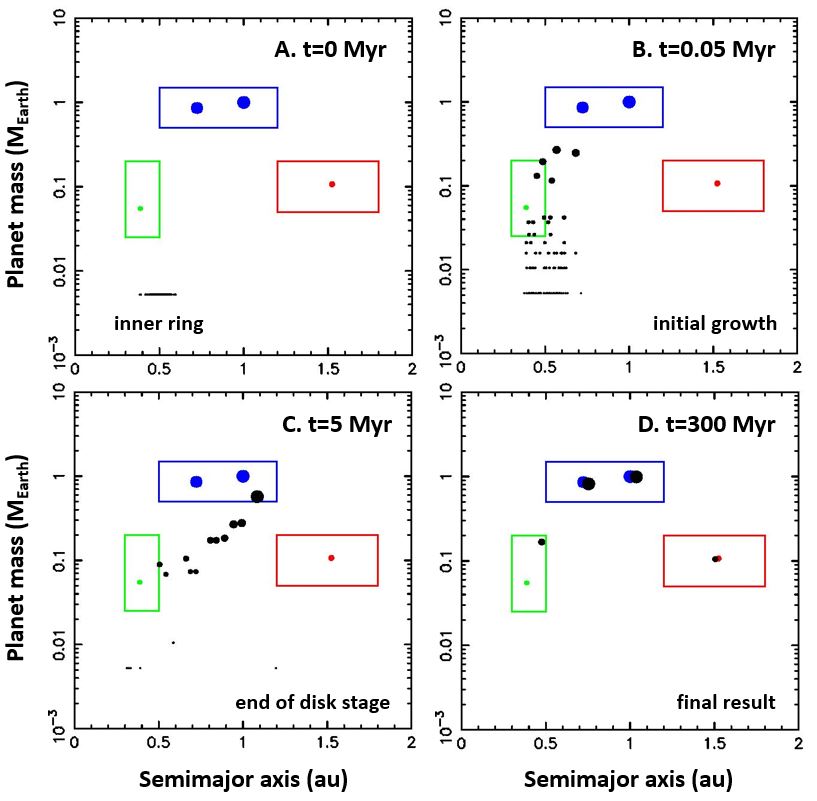}
\caption{The planet growth and migration in the {\tt ring05} model: A. The initial distribution of planetesimals in the
  inner ring with $r_1=0.5$ au and $\sigma_1=0.05$ au. B. The growth of protoplanets from the inner ring. C. The end of
  the gas disk stage. Several relatively large protoplanets formed and migrated to $\sim 1$ au. D. The final result of
  the simulation with four good terrestrial planets. The rectangles delimit our target regions for good terrestrial
  planets (Section 3.1). The colored dots show the semimajor axes and masses of the real planets. The figure illustrates
  a rare case where good Mars formed in the {\tt ring05} simulation -- only 1.3\% of simulations produced good planets,
  including Mars, in this model.}
\label{ringevol}
\end{figure}

Figure \ref{ringevol} shows planet growth and migration in one simulation using an inner planetesimal ring
({\tt ring05} in Table 1; $r_1=0.5$ au and $\sigma_1=0.05$ au). Here we experimented with different migration
rates and found that the results with larger initial gas densities are generally better. We therefore used
$\Sigma_0=3000$ g cm$^{-2}$ in {\tt ring05} (and {\tt ring03}), corresponding to a migration rate about
two times faster than in Fig. \ref{mig}. With this setup we obtained a 39\% success rate for Venus/Earth and
31\% success rate or Mercury, results that are similar to the case with $r_1=0.85$ au ({\tt gas1} in Table 1).
Unfortunately, Mars does not form very often in this setup -- only about 8\% of simulations with good
Venus/Earth also give good Mars. Mars would have to form at larger orbital radii, which is something we
do not take into account in the inner-ring model (see Section 4.8).

Interestingly, the results for the Venus/Earth separation are much better than previous runs, with 44\% (!) 
of the runs with good planets yielding $\Delta a < 0.3$ au. This is a consequence of both the rapid growth of
protoplanets in the inner ring, where the accretion timescales are relatively short, and the outward
migration of the protoplanets. Once the outermost of these planets arrives near the zero-torque radius
at $\sim 1$ au, it stops, allowing the inner planets to catch up.  In turn, this leads to a relatively
strong radial concentration of mass, especially when we use higher gas densities to encourage faster migration.

We tested different ring locations ($r_1=0.3$-0.7 au) and different MDW disks, both with and without turbulent
forcing, to see how the results discussed above depend on these parameters. We identified some parameter
choices that could offer significant benefits over the model discussed above. For example, a model with
$r_1=0.5$ au, $\sigma_1=0.05$ au, $\Sigma_0=3000$ g cm$^{-2}$, $\kappa=3\times10^{-6}$, and $w_3=0.2$
(i.e., 20\% of planetesimals in the radially extended background), yields (V/E, Mars, Merc, V/E Sep) $=$
(49\%, 13\%, 43\%, 55\%) ({\tt back05} in Table 1). These values represent the highest overall success rates
found so far. It shows that it is possible to improve the results for Mercury by including a background
component (from 8\% without background to 13\% with background). The results without turbulence ($\kappa=0$)
were only slightly inferior to those mentioned above. 

In summary, when the formation of planetesimals in the inner ring is combined with strong outward migration,
the Venus/Earth separation becomes significantly better, with up to a 55\% success probability (Table 1) -- a huge
improvement over the previous models. With the inner ring located at $\lesssim 0.5$ au, a good Mercury
forms in 31-56\% of simulations with a good Venus/Earth, which is very promising as well. On the downside,
it is difficult to produce a good Mars in the inner-ring model unless the planetesimal disk extension
beyond 1 au is included in the model. For a more complete picture of the terrestrial planet formation in
a gas disk, we must therefore consider planetesimals/protoplanets that formed and accreted beyond 1 au.
This is the subject of the next few sections.       

\subsection{Extended planetesimal disks and convergent migration}

Here we investigated models with radially extended initial distributions of planetesimals. Two different approaches 
were adopted: (1) we set $w_3=1$ (ignored the ring in Eq. (\ref{weight})) and distributed the background planetesimals
in a disk between $r_{\rm min}$ and $r_{\rm max}$ (Eq. \ref{back}), and (2) we set $w_3=0$ (ignored the background in Eq.
(\ref{weight})) and increased the value of $\sigma_1$ ($\gtrsim 0.2$ au) such that the initial planetesimal
ring was extended over a relatively large range of orbital radii. In these models the convergent migration is the
only physical effect that can concentrate mass near 0.7-1 au. 

For (1), we explored different disk profiles with $-5 \leq \gamma \leq 0$ and varied $r_{\rm min}$ and
$r_{\rm max}$ as well. Some of the best results were obtained with $\gamma=-2$, $r_{\rm min}=0.4$ au and
$r_{\rm max}=1.7$~au, and the standard MDW disk with convergent migration (Figs. \ref{prof} and
\ref{mig}). We obtained (V/E, Mars, Merc, V/E Sep) $=$ (49\%, 12\%, 19\%, 11\%) ({\tt power1} in
Table 1). The success probabilities for Mars and the Venus/Earth separation are relatively low. 
Note that we also tried to decrease the Venus/Earth separation by enhancing convergent migration with
$\Sigma_0=5000$ g cm$^{-2}$ ({\tt power2} in Table 1). This change slightly improved our results for $\Delta a$,
from 11\% in {\tt power1} to 14\% in {\tt power2}, but the overall probabilities remained low. 

In additional models, more planetesimals were placed in the Mars region to increase the probability of obtaining
a good Mars. We tested flatter background planetesimal profiles with $\gamma=-1$ and $\gamma=0$. For
example, with $\gamma=0$ and $\Sigma_0=5000$ g cm$^{-2}$, we obtained the success rate of 12\% for Mars ({\tt power3}
in Table 1). This value is comparable to the models discussed above. Changing the radial profile of the planetesimal disk 
therefore does not improve the success rate for Mars. 

We note that it is difficult in the extended planetesimals disk models to simultaneously form Mars at $\sim 1.5$
au and produce a tight orbital separation for Venus and Earth. To decrease the orbital separation of
Venus and Earth, one might attempt to impose stronger convergent migration, but this means that Mars-size
protoplanets forming at $\gtrsim 1.5$ au quickly migrate to the zero torque radius (i.e., they do not remain near
1.5 au). Imposing weaker convergent migration has the opposite effect. 

The models with wide planetesimal rings were also unsuccessful (item (2) above). We explored $\sigma_1=0.2$, 0.3
and 0.35 au, $r_1=0.85$, 1 and 1.1 au, and $\Sigma_0=600$, 1500 and 3000 g cm$^{-2}$.
These models suffer from the same problems discussed for the power-law profiles above. For example, the model
with $\sigma_1=0.35$ au, $r_1=1.1$ au and $\Sigma_0=3000$ g cm$^{-2}$ yields (V/E, Mars, Merc, V/E Sep) $=$
(37\%, 15\%, 43\%, 4.5\%) ({\tt hump1} in Table 1). As before, the probabilities for Mars and $\Delta a$ are
low.  We were unable to find a way to improve them by tweaking the initial distribution and migration parameters.
We conclude that the models with extended planetesimal disks, even in the presence of strong convergent
migration (Bro\v{z} et al. 2021), do not work. 

\subsection{Models with two source reservoirs}

Finally, we consider models with two planetesimal reservoirs. The motivation for the inner ring at 0.3-0.7 au comes
from the results of Morbidelli et al. (2022) and Marschall \& Morbidelli (2023) who pointed out that planetesimals
could have formed in a narrow ring near the silicate sublimation line.  As we described in Section 4.6, the
models with inner rings produce solid success for Venus/Earth, and excellent success for Mercury and $\Delta a$.
These models, however, were sub-optimal for Mars (8\% success in {\tt ring05}). It is apparently difficult to form good Mars
from the inner ring. In addition, the isotopic composition of Mars differs from that of the Earth (Section 3.2.3).
This constraint shows the need for additional source reservoirs (other than the inner ring).

We used two source reservoirs, located at 0.3-0.7 au (the inner ring) and 1.2-2.0 au (outer source), and explored
the effect of different parameters on the results. Specifically, we varied each source's location, its radial
extension, the partition of mass between the two sources, and gas disk parameters influencing migration. 

In our reference model ({\tt model203} in Table~2), the inner ring had $r_1=0.6$ au and $\sigma_1=0.05$ au, and the
outer source had a Gaussian distribution with $r_2=1.7$ au and $\sigma_2=0.1$ au. The initial mass was partitioned
between the two rings such that there is the mass $m_1=1.4$ $M_{\rm Earth}$ in the inner ring and the mass
$m_2=0.7$ $M_{\rm Earth}$ in the outer source ($w_2=1/3$ in Eq. 2). As for the disk and
migration parameters, we had $r_0=0.9$ au and $\Sigma_0=3000$ g cm$^{-2}$. The effects of turbulence were included
with $\kappa=3 \times 10^{-6}$ and $\tau_\kappa \sim P_{\rm orb}$, where $P_{\rm orb}$ is the orbital period
(turbulence modestly improves the results; see a brief discussion in Section 5). We removed
the disk with one e-fold $\tau=1$ Myr; the migration, damping and turbulent torques exponentially decreased on this
timescale.\footnote{We found that the results are relatively insensitive to the assumed timescale of disk removal.
  For example, if the disk is removed on a shorter or longer timescale, say $\tau = 0.5$~Myr or
  $\tau = 2$ Myr, it is possible to adjust the initial gas density such that the planets grow and migrate
  in much the same way as with $\tau = 1$ Myr. We do not quantify this degeneracy in detail within this paper.}  

\begin{figure}
\epsscale{0.8}
\plotone{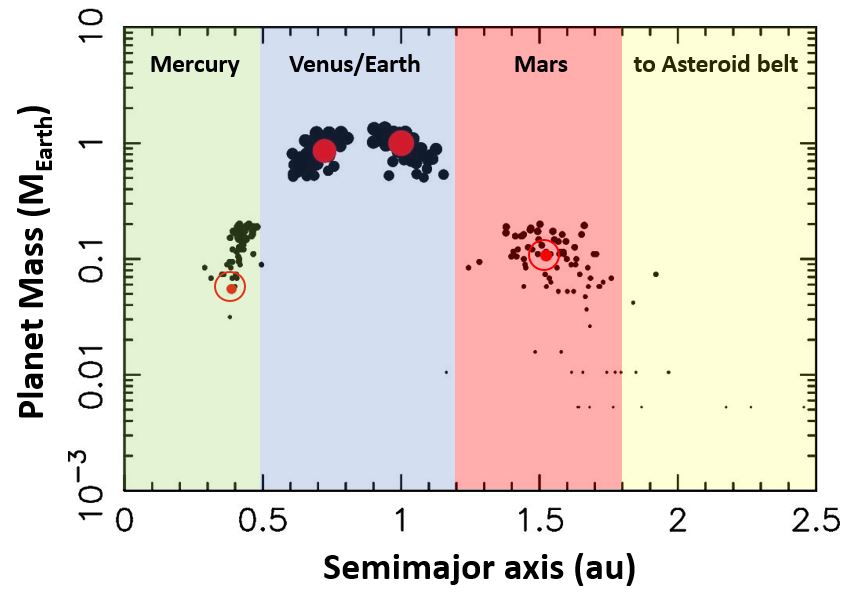}
\caption{The result of 61 simulations from {\tt model203} ($r_1=0.6$ au and $r_2=1.7$ au, no background, strong convergent
  migration) that simultaneously produced a good match to Mercury, Venus, Earth and Mars (as defined in Section 3.1). The black
  dots show the final results in each simulation. The bigger red dots are the real planets.}
\label{plot1}
\end{figure}

\begin{figure}
\epsscale{0.7}
\plotone{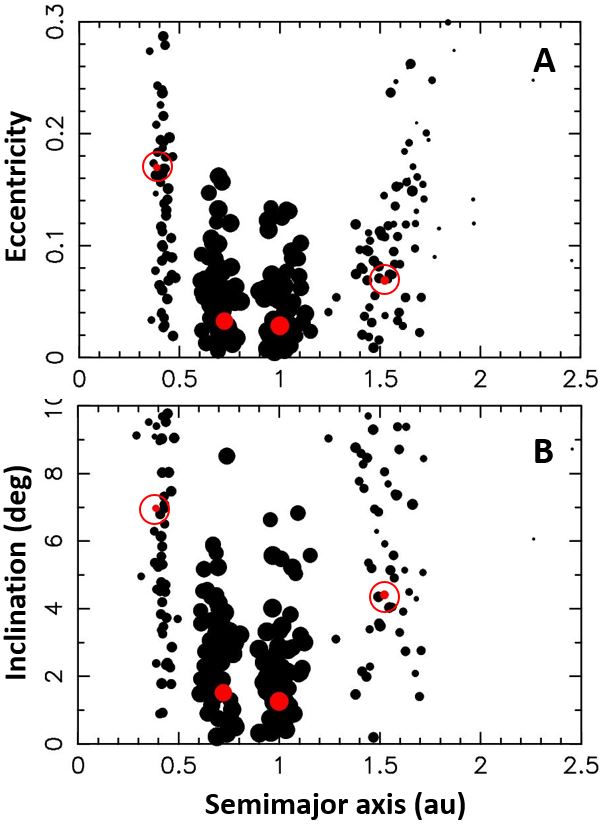}
\caption{The result of 61 simulations from {\tt model203} ($r_1=0.6$ au and $r_2=1.7$ au, no background, strong convergent
  migration) that simultaneously produced a good match to Mercury, Venus, Earth and Mars (as defined in Section 3.1). The black
  dots show the final results in each simulation. The bigger red dots are the real planets.}
\label{plot2}
\end{figure}

\begin{figure}
\epsscale{0.7}
\plotone{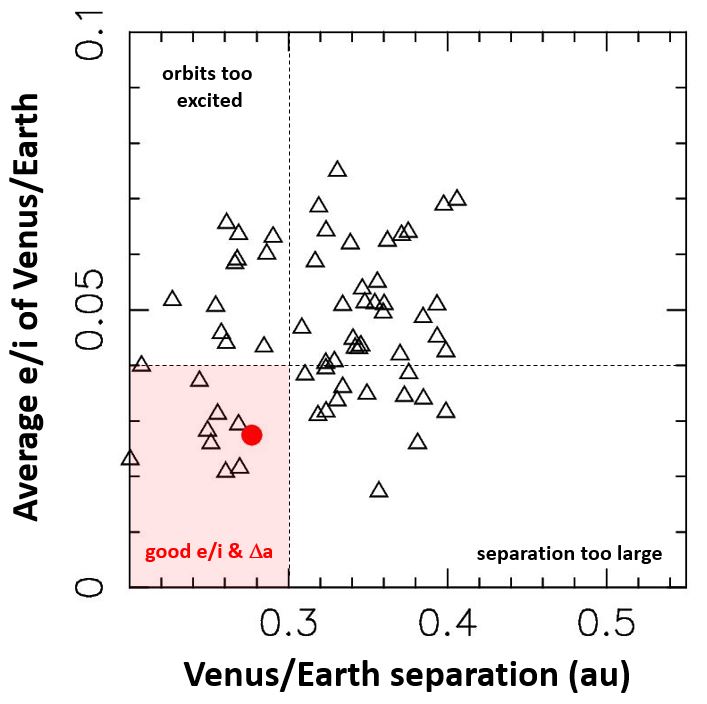}
\caption{The Venus/Earth separation, $\Delta a$, and the average eccentricities/inclinations, $\langle e,i \rangle$,
  for 61 simulations from {\tt model203} ($r_1=0.6$ au and $r_2=1.7$ au, no background, strong convergent
  migration) that produced good terrestrial planets (as defined in Section 3.1). The triangles show the
  results of individual simulations with good final planets. The red dot is the real planets.  We found that
  57\% of simulations produced good AMD of the terrestrial planet system with $S_{\rm d}<0.005$ (Eq. 11; to be compared to
  real $S_{\rm d}\simeq0.003$.}
\label{plot3}
\end{figure}

The reference model ({\tt model203} in Table 2) gives some of the best results obtained so far: (V/E, Mars, Merc,
V/E Sep) = (50\%, 35\%, 32\%, 33\%) (Fig.~\ref{plot1}). The overall success of
{\tt model203}, including all good planets and no bad planets, is 6.1\% (i.e., 61 of our 1000 simulations
satisfied all criteria). The orbital excitation of planetary orbits looks good (Fig. \ref{plot2}; e.g., 57\% of
orbits have $S_d<0.005$; Section 3.1). Notably, 33\% of simulations with good planets have $\Delta a < 0.3$ au
(Fig. \ref{plot1}).\footnote{If we ignore the Mercury constraint and evaluate $\Delta a$ for the simulations 
that produced good Venus/Earth/Mars, we find that 26\% of these simulations have $\Delta a < 0.3$ au.
If we ignore the Mars constraint and evaluate $\Delta a$ for the simulations that produced good Mercury/Venus/Mars, 
we find that 25\% of these simulations have $\Delta a < 0.3$ au. If we ignore both Mercury and Mars constraints
and evaluate $\Delta a$ for the simulations that produced good Venus/Earth, we find that 19\% of these simulations 
have $\Delta a < 0.3$ au. This means that there is a correlation between satisfying the Mercury and/or Mars 
constraints on one hand and having good $\Delta a$ on the other hand. The largest success probability for $\Delta 
a < 0.3$ au occurs in the simulations that end up with four good planets.}
Figure \ref{plot3} shows the distribution of $\Delta a$ and $\langle e,i \rangle$ for the
successful simulations. This figure can be compared with Fig. \ref{delta}, demonstrating that the RMC problem is
alleviated when the simulations account for the convergent migration of protoplanets in a protoplanetary disk.
The total mass of model Earth and model Venus, when averaged over all simulations with good planets, is 1.76
$M_{\rm Earth}$, very close to the real value (1.82 $M_{\rm Earth}$). The mean orbital radius of Venus and Earth,
averaged over all simulations with good planets, is 0.85 au, again very close to the real value (0.86 au).

We explored different model parameters by simulating dozens additional models (Table~2). Some of these models
lead to an improvement over the reference model, at least in some success criteria, while many do not. Notably,
we found higher success for Mercury when the inner ring was placed at $r_1=0.5$ au, or even 0.4 au, than
at $r_1=0.6$ au or 0.7 au. For example, {\tt model217} with $r_1=0.5$ au gives a good Mercury in 40\% of
simulations in which a good Venus/Earth form, whereas {\tt model222} with $r_1=0.4$ au yields a good
Mercury in 58\% of all simulations. The model with $r_1=0.5$ au, however, provides a better orbital radius
for Mercury than the model with $r_1=0.4$ au (in which Mercury ends up some 0.05-0.1 au inward of its real location). 
This trend continues and we get a 65\% success for Mercury with $r_1=0.3$ in {\tt model223}, but Mercury typically
ends up some 0.1-0.2 au inward of its real location. In contrast, Mercury's orbital radius is roughly correct
in the models with $r_1=0.5$ or $r_1=0.6$ au. These cases therefore offer the best benefits with both the
relatively high formation probabilities and correct orbital radii of Mercury analogs.

The models with more distant inner rings were less successful (e.g., only a 22\% success for Mercury in
{\tt model218} with $r_1=0.7$ au). With $r_1=0.7$ au, there is apparently not enough mass below 0.5 au for a good
Mercury analog to form. Also, scattering bodies from $>0.5$ au to $<0.5$ au has a low efficiency due to the
damping effects of the gas disk. We conclude that $r_1\simeq0.5$-0.6 au is the optimal location of the inner
ring.

It does not seem to matter for the results if the inner ring was narrow ($\sigma_1 = 0.05$ au, about 10\% of
its heliocentric distance), very narrow ($\sigma_1 = 0.025$ au, about 5\%), or slightly wider ($\sigma_1 = 0.1$ au,
about 20\%). All these models give similar success probabilities for Mercury (36-44\% in {\tt model225},
{\tt model217} and {\tt model224}), and do not differ in other criteria as well. The inner ring cannot be wide,
however, because {\tt model235} with $\sigma_1 \simeq 0.2$ au fails for Mercury (only a 15\% success) and
the radial separation of Venus and Earth (zero success). The accretion timescale is longer in a wide ring
and this reduces the planet growth and limits the effects of convergent migration. We conclude that the inner
ring was relatively narrow ($\sigma_1 \lesssim 0.1$ au), which is consistent with the idea of early
planetesimal formation at the silicate sublimation line (Morbidelli et al. 2022, Marschall \& Morbidelli 2023). 

We find that the outer source can be radially extended and located as far as $\sim 2$~au, or possibly even beyond.
Specifically, when we set $\sigma_2=0.2$ au, 0.3 au or 0.4 au, instead of $\sigma_2=0.1$ au as in the
reference model, the results are similar to the reference model (compare {\tt model219}, 
{\tt model227} and {\tt model234} with {\tt model203} in Table 2). The success rate for Mars increases with
$\sigma_2$, from 28\% for $\sigma_2=0.1$ au in {\tt model203} to 39\% for $\sigma_2=0.4$ au in {\tt model234},
indicating that {\it wider} outer reservoirs would work better for Mars formation. We thus find that the outer
planetesimals were probably not distributed in a narrow ring (like the inner planetesimals) but were instead
extended over a range of orbital radii. Some concentration of planetesimals beyond 1.5 au is probably needed,
however, because the {\tt back05} model from Section 4.6 -- a combination of the inner ring at 0.5 au with
a power-law distribution of background planetesimals -- only showed a 13\% success for Mars.

The outer source with $r_2=1.9$ au or $r_2=2.1$ au works slightly better for Mars (38\% success rate in
{\tt model215} and 37\% success rate in {\tt model229}) than $r_2=1.7$ au (35\%; {\tt model203}), but this
small difference is probably not significant. Interestingly, the success rate for Mars drops to 21\% in a model
with $r_2=1.5$ au ({\tt model216} in Table 2). This would suggest that the outer reservoir was located
beyond 1.5 au. 

The strength of convergent migration matters as well. The best results for the two-source model were obtained
with $\Sigma_0=3000$ g cm$^{-2}$. The radial separation of Venus and Earth increases for $\Sigma_0=1500$ g
cm$^{-2}$ (only a 16\% success for $\Delta a < 0.3$ au in {\tt model 213}) and decreases for $\Sigma_0=6000$ g
cm$^{-2}$ (a 47\% success for $\Delta a < 0.3$ au in {\tt model 214}). This makes sense because $\Sigma_0$
controls the strength of convergent migration and must influence the radial separation of Venus and Earth.
The opposite effect is found for Mars when varying $\Sigma_0$. With $\Sigma_0=6000$ g cm$^{-2}$, a good Mars
forms in only 20\% of simulations (to be compared to 34\% in the reference model). This result is to be
expected: if the migration is stronger, protoplanets move away from the outer source reservoir, leaving an
insufficient mass to form a good Mars. As for Mars, the results with $\Sigma_0=1500$ g cm$^{-2}$ are
similar to those with $\Sigma_0=3000$ g cm$^{-2}$.  This shows that decreasing the migration rate below
that of the reference model is inconsequential for Mars. We conclude that $\Sigma_0 \simeq 3000$ g cm$^{-2}$
is a good compromise between obtaining reasonable results for Mars and finding the right radial separation
for Venus and Earth.

The results with increased resolution, $N=1000$ and $N=2000$, confirm the findings reported above. For example,
for the reference model and $N=1000$, we obtain (V/E, Mars, Merc, V/E Sep) $=$ (47\%, 32\%, 32\%, 27\%),
({\tt model231} in Table 2), which can be compared to (V/E, Mars, Merc, V/E Sep) $=$ (50\%, 35\%, 32\%,
33\%) in {\tt model203} with the standard resolution. The results with $N=2000$ are also similar. It would be
desirable to increase the resolution further, possibly employing the GENGA code (Grimm \& Stadel 2014) on
GPUs, to see if any interesting differences could be identified with $N \sim 10^4$.  

\subsection{Planet accretion in the two-source models}

\begin{figure}
\epsscale{0.7}
\plotone{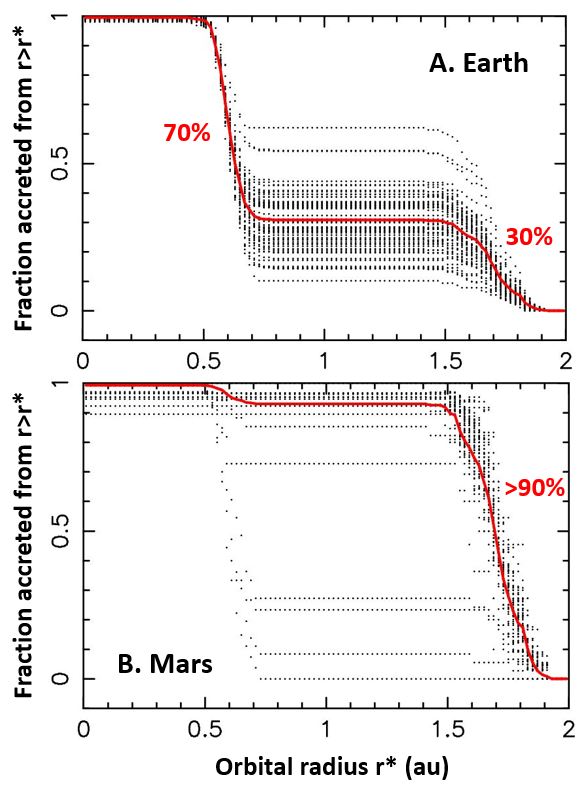}
\caption{The fraction of Earth and Mars mass accreted from beyond the orbital radius $r^*$, as a function of $r^*$.
  We collected all simulations in {\tt model203} ($r_1=0.6$ au and $r_2=1.7$~au, the 2:1 mass split between the inner
  and outer sources, no background, strong convergent
  migration) that produced good matches to the terrestrial planet system (i.e., four good planets as defined
  in Section 3.1). We then followed the accretion sequence for Earth and Mars and collected all
  planetesimals that accreted into Earth and Mars, respectively. This accounts for all accretion 
  events, including planetesimals accreting into intermediate-mass protoplanets which then merged 
  into the identified Earth/Mars analogs. The cumulative distribution Earth's and Mars' fractional mass budgets,
  $f(r>r^*)$, is  plotted here. The black dots represent the results of individual simulations. The red line 
  is the average. On average, the Earth accreted $\simeq 30$\% of its mass from the outer reservoir.}
\label{plot5}
\end{figure}

The source of material accreted by Earth and Mars in {\tt model203} is shown in Fig.~\ref{plot5}.
For $w_2=1/3$ (i.e., the 2:1 mass split between the inner and outer sources), we
find that the Earth accretes about 70\% of its terminal mass from the inner ring and about 30\% of its mass from
the outer ring. The inner ring dominates the early accretion (Fig.~\ref{ringevol}). The outer source material
is added later when the proto-Earth reaches $\sim 1$ au and starts accreting the material that migrated
to $\sim 1$ au from the outer source.
Assuming that the inner ring material was reduced and the outer ring material was oxidized,
as expected from temperature profile in the protoplanetary disk, these results would be consistent with the
elemental composition of the Earth (i.e., the early accretion of $\sim 70$\% reduced material followed
by the later accretion of $\sim 30$\% oxidized material; Section 3.2.1) (Rubie et al. 2011, Dale et al. 2025). 

In all these models, Mars accretes 90\% of its materials, on average, from the outer source (Fig. \ref{plot5}).
This would be consistent with the isotopic composition of Mars, which is distinct from that of the Earth,
and with the more oxidized nature of Mars' mantle (Section 3.2.3). These constraints would be more difficult
to satisfy in a single source model (e.g., Sections 4.5 and 4.6).

\begin{figure}
\epsscale{0.6}
\plotone{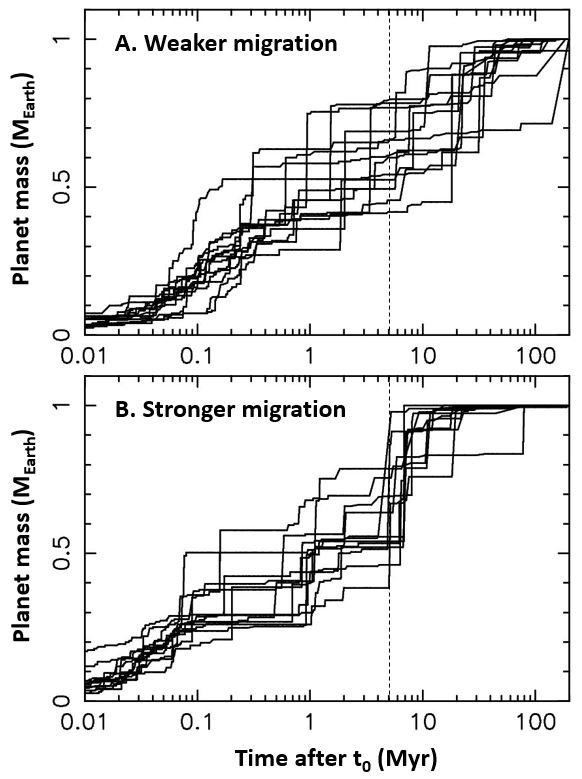}
\caption{Earth's growth in the successful simulations from {\tt model213} (panel A, weaker migration with
  $\Sigma_0=1500$ g cm$^{-2}$) and {\tt model214} (panel B, stronger migration with $\Sigma_0=6000$ g cm$^{-2}$).
  We collected all successful simulations that produced good analogs to the terrestrial planet
  system (four planets with good orbits and masses; Sect. 3.1). In addition, we only used those simulations
  in which the final mass of the Earth analog was withing 10\% to the real Earth mass. The accretion sequence
  of these very good Earth analogs was recalibrated such that the model Earth ended up exactly with
  1~$M_{\rm Earth}$. The vertical dashed line marks the end of the gas disk stage.}
\label{delayed}
\end{figure}

We find that the nature of Earth's growth during late stages mainly depends on: the (i) strength of convergent migration
during Stage 1, and (ii) initial location of the outer ring. For strong convergent migration ({\tt model214} with
$\Sigma_0=6000$ g cm$^{-2}$) and/or the outer ring at $r_2=1.5$ au ({\tt model216}), the Earth accretes relatively
quickly after the gas disk dispersal, and there are fewer large accretion events at late times ($t>10$ Myr, panel B.
in Fig. \ref{delayed}). This is a consequence of strong mass packing in these models during Stage 1. 
Many of these cases could be ruled out by the chronology of Earth's accretion inferred from
the Hf/W system (Kleine \& Walker 2017).\footnote{There are exceptions with several individual simulations in
  Fig. \ref{delayed}B showing very long accretion timescales. The chronology constraints could therefore be satisfied
  in these models, but the probability of that happening is low.}
The results could potentially be improved if the migration/instability of the outer planets was delayed (Woo et al.
2024), but we note that the Earth often reached its near-final mass within $\sim 5$ Myr after the gas disk dispersal
($t=5$-10 Myr in Fig. \ref{delayed}B), before the instability happened in our simulations ($t\simeq11.2$ Myr in Fig. \ref{delayed}B).
Delaying the giant planet instability may thus not significantly help to delay the Earth accretion history in these models.

In the models with weaker convergent migration (e.g., {\tt model213} with $\Sigma_0=1500$ g cm$^{-2}$) and/or when
the outer ring is located further out (e.g., $r_2=1.9$ au in {\tt model215}), the Earth accretion is completed over
longer timescales and there are often several late accretion events happening at $t=10$-50 Myr (Figure
\ref{delayed}A). These models would be more compatible with the chronological constraints.

\begin{figure}
\epsscale{0.6}
\plotone{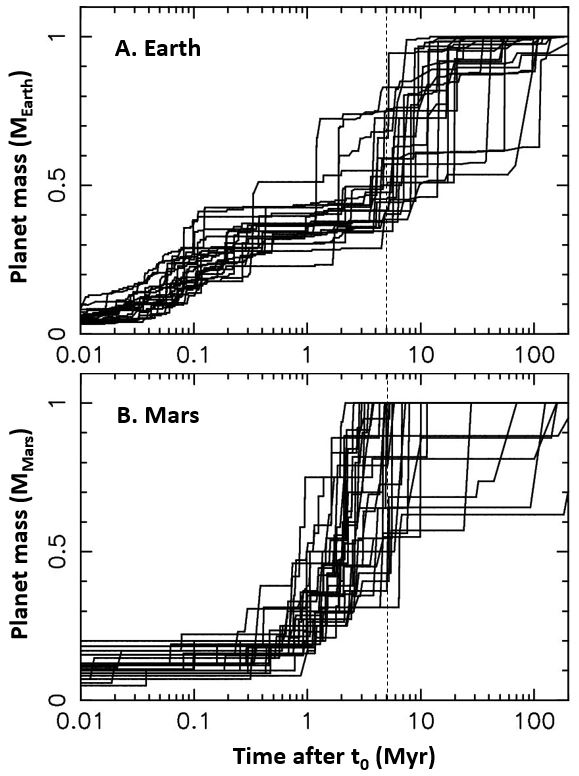}
\caption{The growth of planets in the successful simulations from {\tt model203} (two planetesimal reservoirs
  with $r_1=0.6$ au $r_2=1.7$ au, $\sigma_2=0.1$ au and $w_2=1/3$; the gas disk with $r_0=0.9$ au and
  $\Sigma_0=3000$ g cm$^{-2}$). We collected all successful 
simulations that produced good analogs to the terrestrial planet system (four planets with good orbits 
and masses; Sect. 3.1). In panel A, the accretion sequence in each simulation was recalibrated such that
the model Earth ended up exactly with 1~$M_{\rm Earth}$. In panel B, the accretion sequence was recalibrated such that
the model Mars ended up exactly with 1~$M_{\rm Mars}$. The vertical dashed line marks the end of the disk stage.}
\label{plot4}
\end{figure}

The growth of the Earth in
our reference model ({\tt model203}) is intermediate between the two cases discussed above (Fig. \ref{plot4}A).
The growth of Mars in the reference model is relatively fast: in $\sim 80$\% of successful simulations, the Mars
accretion was practically completed by 10 Myr after time zero (Fig. \ref{plot4}B). 
These cases would be in line with the chronological constraints discussed in Section 3.2.3 (Dauphas \& Pourmand 2011,
Kobayashi \& Dauphas 2013, Marchi et al. 2020, P\'eron \& Mukhopadhyay 2022).

\begin{figure}
\epsscale{0.6}
\plotone{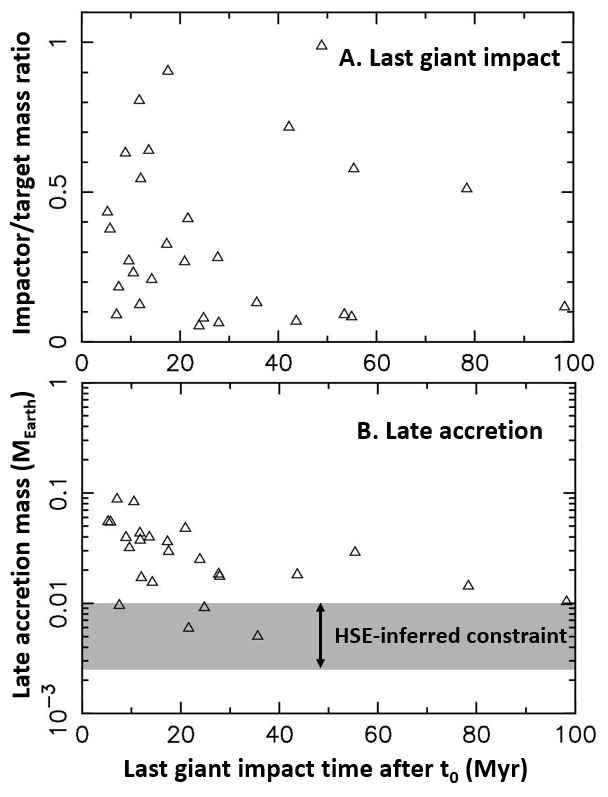}
\caption{Panel {\bf A}. The last giant impact on the Earth in our reference model ({\tt model203}). We collected all
  successful simulations and determined the last giant impact on the proto-Earth (as defined in the main text).
  The triangles show the time of the last giant impact, $t_{\rm giant}$, and the impactor-to-target mass ratio, 
  $\Gamma$. Panel {\bf B}. The mass accreted by the Earth after the last giant impact (aka late accretion). The gray
  band shows the HSE-inferred late accretion, 0.25-1\% $M_{\rm Earth}$, inferred from cosmochemical constraints
  (Morbidelli \& Wood 2015).}
\label{plot6}
\end{figure}

We identified the last giant impact on the Earth in all successful simulations. The giant impact was defined as
an impact on the proto-Earth (i.e., the planet that becomes a good Earth analog at the end of the simulation)
with the target mass $>0.5$ $M_{\rm Earth}$ at the time of impact, and the impactor-to-target mass ratio $\Gamma>0.05$.
For comparison, the canonical Moon-forming impact corresponds to $\Gamma \sim 0.1$ (Canup et al. 2023).
Figure \ref{plot6}A shows the properties of the last giant impact on the Earth in {\tt model203}. Many of these
cases would be incompatible with the chronological constraints on the Moon-forming impact (Kleine \& Walker
2017), for example those where the last giant impact happened too early ($t_{\rm giant}<20$ Myr).

In {\tt model203}, we find that $\simeq 50$\% of last giant impacts happen at $t_{\rm giant}>20$ Myr, and
$\simeq 30$\% of last giant impacts happen at $t_{\rm giant}>40$ Myr. The properties of the last giant impact in different models
correlate with the Earth accretion history. For example, the models with stronger convergent migration
({\tt model214} with $\Sigma_0=6000$ g cm$^{-2}$) produce more last giant impacts with $t_{\rm giant}<20$ Myr, whereas
those with weaker convergent migration ({\tt model213} with $\Sigma_0=1500$ g cm$^{-2}$) have more cases with
$t_{\rm giant}>20$ Myr. In all models, the impactor-to-target mass ratio shows a wide range of values ($0.05<\Gamma<1$;
e.g., {\tt model203} in Figure \ref{plot6}A). Our simulations therefore do
not have the predictive power to inform the exact nature of the Moon-forming impact (Canup et al. 2023).\footnote{We find that
  a giant impact with $t_{\rm giant}<20$ Myr and $\Gamma>0.1$ is often followed by an impact with $t_{\rm giant}>20$ Myr and
  $0.01<\Gamma<0.1$. If the first impact is interpreted as the Moon-forming impact, the Moon would form early,
  perhaps too early for the cosmochemical chronometers, but the second impact would have the potential to affect the
  chronometers as well. A thorough investigation of this issue is left for future work.} 

We find that the Moon-forming impactor, Theia, often shares the accretion history the Earth (i.e., grows from
the inner ring on a similar timescale as the Earth). This result suggests that Theia could have had an isotopic
composition similar to that of the Earth. It would help to explain why the Moon and Earth have remarkably similar
isotopic compositions for many elements (e.g., Canup et al. 2023). 

The initial resolution of planetesimal sources with only $N=400$ bodies limits our ability to correctly evaluate issues
related to late accretion on the Earth (Morbidelli \& Wood 2015). Each of our initial planetesimals has the mass
$\simeq 0.5$\% of the Earth mass, which is similar to the HSE-inferred late accretion (Anders 1968, Turekian \&
Clark 1969). This means that a late impact of a single `planetesimal' in our simulations would bring the entire mass
required for HSEs. The results are not expected to be particularly reliable in this situation. Still, for completeness,
Fig. \ref{plot6}B shows the mass accreted by the Earth after the last giant impacts recorded in {\tt model203}
(Fig. \ref{plot6}A). We find that accreted mass is often a factor of several larger than the HSE-inferred late
accretion, especially if $t_{\rm giant}<20$ Myr. This argument would tentatively favor $t_{\rm giant}>20$ Myr (also see
Woo et al. 2024), but the resolution of the planetesimal disk would have to be improved before any stronger
conclusions can be drawn from this. We repeated the analysis of {\tt model203} with improved resolution ($N=1000$
and $N=2000$) and found that the results are nearly identical to those shown in Fig. \ref{plot6}. 
There is also the
possibility that some of the added mass would go to the Earth core, reducing the excess delivered to the mantle (e.g.,
Korenaga \& Marchi 2023).

\subsection{Models with collisional fragmentation}

Our model for collisional fragmentation was described in Section 2.5. The collisional fragmentation is not the main
focus of this work but we nevertheless performed several simulations with fragmentation for completeness. In one of these
tests, we re-run the reference model ({\tt model203}), using exactly the same initial conditions, resolution and
gas disk parameters, but this time accounting for the collisional fragmentation as well ({\tt model233} in Table~2).

The collisional fragmentation model yields (V/E, Mars, Merc, V/E Sep) $=$ (31\%, 36\%, 32\%, 41\%), which can be compared to
(V/E, Mars, Merc, V/E Sep) $=$ (50\%, 35\%, 32\%, 33\%) in the reference model without fragmentation. We find that
the success probability for Venus/Earth somewhat decreased. This is probably a consequence of the mass removal in our
fragmentation algorithm (Section 2.5). With small fragments  being removed in the simulation, the final masses of Venus
and Earth are $\simeq 10$\% smaller than they should be (the average total mass $\simeq 1.6$ $M_{\rm Earth}$ vs.
the real mass 1.82 $M_{\rm Earth}$), and this makes it slightly more difficult to satisfy our condition for
good Venus/Earth (Section 3.1.1). This problem could we resolved if we used a slightly larger total mass of
the initial planetesimals (e.g., 2.3-2.4 $M_{\rm Earth}$ instead of the original mass 2.1 $M_{\rm Earth}$). Alternatively, our
collisional fragmentation algorithm exaggerates mass wasting.

We also noted several additional differences and similarities between the simulations with and without fragmentation. The
Venus/Earth separation is slightly better with fragmentation (41\% in {\tt model233}) than without it (33\% in {\tt model203}),
but this difference is not necessarily significant, because it can be influenced by the problem with the Venus/Earth
masses discussed above. As the Venus/Earth masses in {\tt model233} ended up to be smaller than they should be,
this problem could have affected the orbital separation by allowing the two (lighter) planets to have orbits
that are closer together.

The orbital excitation of planetary orbits does not change with fragmentation (the success rate for $S_{\rm d}<0.005$ remains
near 50\%). This result, however, does not mean that the collisional fragmentation is unimportant for AMD, because the results
can be influenced by our fragmentation algorithm where we do not keep small fragments in simulations. If the small fragments
were included, they could exert dynamical friction on larger protoplanets and potentially reduce their orbital excitation.
Note that the previous work on this issue did not found any major influence, as for AMD, in the models with and without collisional
fragmentation (e.g., Chambers 2013, Woo et al. 2024). The AMD problem can be more fundamentally affected by the initial
resolution of terrestrial planetesimals in the $N$-body code (Section 4.9; O'Brien et al. 2008). Indeed, our simulation
of {\tt model203} with $N=2000$, and no collisional fragmentation, gives a 75\% probability to have $S_{\rm d}<0.005$. 

The success rate for Mercury does not change but Mercury's mass is slightly smaller in the simulations with fragmentation
(the average mass of good Mercury analogs is $\simeq 0.09$ $M_{\rm Earth}$ vs. the original $\simeq 0.13$ $M_{\rm Earth}$,
to be compared to real $M_{\rm Mercury}=0.055$ $M_{\rm Earth}$). This difference, about 30\% in terms of Mercury's mass, is more
significant than the change of the Venus/Earth mass, and it would probably not be wiped out if the initial mass of
planetesimals was increased by 10\% (as discussed above). It is a consequence of the stronger orbital excitation and larger
impact speeds of protoplanets in the Mercury's formation zone ($\sim 0.4$ au). When we adopt a strict definition of good Mercury
analogs, $0.025<m<0.1$ $M_{\rm Earth}$ instead of the standard $0.025 < m <0.2$ $M_{\rm Earth}$, the success rate with this new
definition improves from 9\% in {\tt model203} (with perfect mergers) to 17\% in {\tt model233} (with collisional fragmentation).

As we discussed in Section 3.2.4, Mercury has a large metallic core representing $\simeq$ 70\% of Mercury's mass (Hauck
et al. 2013; the Core Mass Fraction or CMF $\simeq 0.7$). The effect of hit-and-run and disruptive collisions in our
{\tt model233} is probably not large enough to fully explain Mercury's CMF. For example, if we set a reference CMF = 1/3,
similar to the Earth (Poirier 2000), and assume that $\sim 30$\% of Mercury's mass was removed by impacts  
(see above; all mantle material), we find that the CMF would (on average) increase to $\simeq 0.6$. This result neglects numerous 
small mantle fragments that were removed in our simulations, but could have in reality accreted back on Mercury, if they 
were not removed. We therefore infer CMF $< 0.6$. These findings agree with those of Scora et al. (2024) who found 
only a modest change in Mercury's CMF with fragmentation. {Clement et al. (2023) performed a similar analysis and 
obtained CMF = 0.6-0.8 in 10-20\% of their simulations that produced a reasonable Mercury analog. While these results
suggest that the mantle removal during collisions can help to increase Mercury's CMF, the probability of starting with 
CMF = 1/3 end ending with CMF $\sim 0.7$ is found to be somewhat low. Disruptive collisions of planetesimals\footnote{A better 
$N$-body treatment for collisional fragmentation of planetesimals will need to be developed to test this. The planetesimal 
stage is ignored in our current fragmentation code.}, sublimation of planetesimal silicates (Eriksson et al. 2021), and /or 
formation of planetesimals from iron-rich pebbles (Johansen \& Dorn 2022) could contribute to increasing Mercury's CMF. 
We will address Mercury's CMF in future work.}

\section{Discussion}


Several cosmochemical constraints provide support for our two-source model of terrestrial planet accretion.  
Rubie et al. (2011) and Dale et al. (2025) modeled the chemical composition of the Earth mantle. These works showed that
it is difficult to simultaneously match the SiO$_2$ and FeO concentration in the BSE if the Earth accreted from
a uniformly reduced or uniformly oxidized reservoir. They suggested that this problem could be solved if the Earth
initially accreted $\sim 70$\% of material from a reduced reservoir (similar to Enstatite Chondrites in oxidation but
enriched in refractories and s-isotope products), and subsequently accreted $\sim 30$\% of material from an oxidized
reservoir (similar to Ordinary Chondrites or non-carbonaceous iron meteorite parent bodies; Grewal et al. 2025). 
The inner reservoir is not expected to be sampled in the current meteorite collection (see below).
This agrees with the accretion sequence of the Earth in the two source model because
the Earth initially grows by accreting the material from the inner ring, which should be reduced given higher
disk temperatures at $\sim 0.5$ au, and later incorporates a smaller amount of material from the outer source, which is
expected to be more oxidized.

For the 2:1 mass split between the inner and outer sources ($w_2=1/3$ in Eq. 2), we find that the Earth
accretes $\sim 70$\% of material from the inner ring and $\sim 30$\% of material from the outer source (Fig. \ref{plot5}A).
This case would represent the best match to the chemical composition of the BSE (Rubie et al. 2011, Dale et al. 2025).

We found that the inner ring reservoir should be unsampled in the current meteorite collection. For that, we
followed the orbital evolution of every inner ring planetesimal and checked that none of these planetesimals ended
on a stable orbit in the asteroid belt, where it could be disrupted and produce meteorites. This does not represent a
strong constrain when we consider {\it one} of our simulations because we only had $\sim 270$ inner-ring planetesimals
with $w_2=1/3$ in models listed in Table 2 (the models with the increased resolution had $\sim 670$ inner ring
planetesimals for $N=1000$ and $\sim 1330$ for $N=2000$). Collectively, however, we have $\sim 500$ simulations
available in each model with good Venus and Earth. Given that none of the inner ring planetesimals ended in the asteroid
belt in these simulations we can estimate that the implantation efficiency is $<1/(500\times270)\sim 7\times 10^{-6}$.
Furthermore, given that we investigated $\sim 30$ models with the inner planetesimal ring, the (combined) implantation
efficiency is $\lesssim 3 \times 10^{-7}$. If the total inner ring mass (1.4 $M_{\rm Earth}$) is distributed among diameter
$D=100$ km planetesimals, we estimate that there would be $\sim 3 \times 10^6$ inner-ring planetesimals to start with.
This implies that $<1$ planetesimals would be implanted in the asteroid belt. For comparison, there are over
200 main belt asteroids with $D>100$ km. This shows that the inner ring material should represent a negligible source of
meteorites.

The second line of support for the two-source model comes from the isotopic composition of Mars (Dauphas et al. 2024).
The composition of Bulk Silicate Mars (BSM) in terms of moderately siderophile elements like chromium and titanium
is more similar to enstatite chondrites (EC), but in terms of highly siderophile elements like molybdenum, Mars
is more similar to ordinary chondrites (OC). Ignoring the possibility of accretion from an unsampled reservoir,
Dauphas et al. (2024) used this argument to argue that Mars most likely accreted $\sim 2/3$ of its mass form the EC
reservoir and $\sim 1/3$ from the OC material.

Interestingly, the OC material accretion by Mars is inferred to predate the EC accretion (Dauphas et al. 2024). If we
tentatively associate the EC material with the inner ring and OC material with the outer ring, we would potentially
be able to explain this constraint because Mars would first accrete OC material from the outer ring and then EC
material from the inner ring.
On the other hand, a 2:1 mass split between the inner and outer sources would produce a Mars that accretes $\sim 90$\%
of material from the outer source and $\sim 10$\% of material from the inner ring (on average; Fig. \ref{plot5}B).
This composition would not have enough EC material to explain observations. We suspect the resolution of this issue
lies in the exact radial distribution of unsampled, EC and OC materials in the protoplanetary disk. 

Jacobson et al. (2017) hypothesized that the accretion of Venus was characterized by the absence of late giant impacts
and the preservation of its primordial stratification (Section 3.2.2). In the classical model of late stage accretion
(Chambers 2001, Hansen 2009), in which several Mars-class protoplanets survive near 1 au, 
it is difficult to avoid late giant impacts on Venus. The chances of this happening increase in the model where {\it fewer}
larger protoplanets form near 1 au, because fewer late giant impacts happen in this case. It then becomes more likely
to have the Moon-forming impact on the Earth and no late giant impact on Venus. The Venus/Earth formation from the inner 
inner ring leads to fewer larger protoplanets at the end of the gas disk stage and provides a possible justification
for this hypothesis. 

In many models investigated here, we often recorded large impacts on Mars happening $\sim 200$ Myr after $t_0$. Some of the
potential impactors even survive in the 1.5-2 au region at $t=300$ Myr (Fig. \ref{plot1}). This can provide some
justification to the suggestion of Morbidelli et al. (2018) that the Mars impact record requires a global resurfacing
event, probably related to the formation of the Borealis basin (Marinova et al. 2008, Nimmo et al. 2008), at
$\sim 4.4$ Gyr ago. In this context, the Borealis impact would be the last major impact on Mars and the
one that created the North-South topographic dichotomy. Previous large impacts would also happen, but they would not
leave any obvious topographic signatures (e.g., Andrews-Hanna \& Bottke 2017). Smaller impactors from the $\sim 1.5$~au
region would also contribute to the formation of lunar basins, including Imbrium (Nesvorn\'y et al. 2023).

Some level of turbulence increases the chances of good Mercury to form. For example, in the reference
model ({\tt model203}) with turbulence ($\kappa=3 \times 10^{-6}$ and $\tau_\kappa \sim P_{\rm orb}$), we get good Mercury in
32\% of cases. When we switch the turbulence off, the success rate for Mercury drops to 16\%.
Turbulent forcing helps the protoplanets near 0.5 au to grow and migrate. Without it, small protoplanets isolate themselves.
They do not grow to large enough sizes, do not migrate, and leave too much mass at Mercury's location. 
The level of the turbulence used here would be consistent with $\alpha \sim 10^{-3}$ in the inner disk (Okuzumi \& Ormel 2013),
which is intermediate between our simple disk, where $\alpha = 10^{-4}$ at all orbital radii, and the results of
Marschall \& Morbidelli (2023) who suggested that the viscosity can be as high and $\alpha \sim 0.01$ in the
inner (young) disk. Here we tested and constructed successful migration models for the terrestrial planets with
$\alpha = 10^{-3}$-10$^{-5}$. We therefore believe that different levels of turbulent forcing can apply.
Other sources of stochastic forcing and/or means of breaking protoplanet isolation could potentially be applicable
as well (Armitage 2011). The development of more complete disk models, where the stochastic forcing consistently
varies with the disk properties and evolution, is left for future work. 

Planetesimal formation at the silicate sublimation line, where various effects concentrate solids and trigger
instabilities (Dra{\.z}kowska \& Alibert 2017, Carrera et al. 2021), could provide direct motivation for the
inner planetesimal ring (Morbidelli et al. 2022, Izidoro et al. 2022, Marschall \& Morbidelli et al. 2023),
but see Carrera et al. (2025) who argued that strong turbulence may prevent planetesimal formation in Class 0/I disks.
Note that the formation of planetesimals
in a ring near the silicate condensation line does not require the existence of a pressure bump at this location
(Morbidelli et al. 2022). The origin of the outer source of planetesimals at 1.5-2 au is less clear. For example,
the outer source cannot be associated with the ice line -- the presumed site of formation of carbonaceous
irons and carbonaceous chondrites -- because that would contradict many cosmochemical constraints (e.g., Dauphas
et al. 2024, Morbidelli et al. 2025, Dale et al. 2025). Indeed, in the disk model from Morbidelli et al. (2022),
the ice line is located at $\sim 3$-5 au during the early stages of planetesimal formation. The question of the outer
source origin is deferred to future work (Goldberg et al., in preparation).
  
\section{Conclusions}

We conducted a systematic study of terrestrial planet formation. Our simulations started
  at the protoplanetary disk stage, when planetesimals formed and accreted into protoplanets, and continued
  past the late stage of giant impacts. We explored the effect of different parameters, such as the initial
  radial distribution of planetesimals and Type-I migration of protoplanets, on the final results.
A thousand simulations were completed in each case starting from slightly modified planetesimal
  distributions -- generated with different random deviates -- to characterize the stochastic nature of the
  accretion process. The large number of simulations allowed us to statistically evaluate each model setup
  as for the success probability to produce correct terrestrial planets.

\begin{figure}
\epsscale{0.7}
\plotone{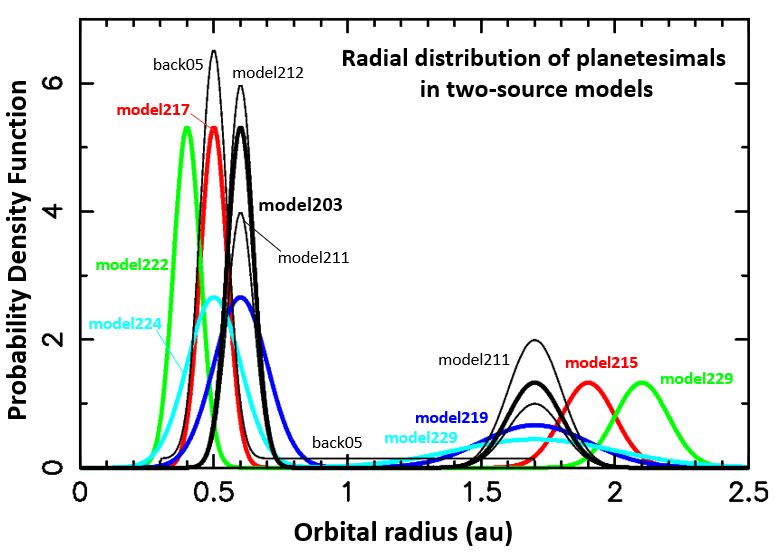}
\caption{The radial distribution of terrestrial planetesimals in the successful two-source models tested in this
  work. To produce correct Mercury, Venus and Earth, the inner ring of planetesimals must have been radially narrow
  ($\sigma_1\lesssim0.1$ au) and was located at $\simeq 0.4$-0.6 au.
  The outer reservoir of planetesimals beyond $\sim 1.5$ au was the primary source of Mars' accretion.}
\label{2source}
\end{figure}

The main conclusions of this work are as follows:
\begin{enumerate}

\item The original model of Hansen (2009), where planetesimals were distributed between 0.7 au and 1 au
  and the gas disk effects were ignored, gives relatively good results for Venus, Earth and Mars but fails to
  produce Mercury and the tight radial separation of Venus and Earth.
\item The two component model, with a planetesimal ring at $\simeq 0.85$ au and 20\% of planetesimals in a
  radially extended background, gives better results for Mercury, but still fails for the Venus/Earth separation.
  When convergent Type-I migration is accounted for in this model, the Venus/Earth separation improves but
  the probability to obtain a good Mercury decreases (as the gas effects limit orbital scattering). 
\item In our best model, the terrestrial planets accreted planetesimals from two sources: the (1) inner
  ring of planetesimals that presumably formed near the silicate sublimation line at $\simeq 0.5$ au
  (Morbidelli et al. 2022, Marschall \& Morbidelli 2023), and (2) outer source at 1.5-2.0 au, which was
  inferred here to be the primary source of Mars accretion. Figure \ref{2source} summarizes the initial
  distribution of terrestrial planetesimals implied by this work. 
\item For Venus and Earth to start in the inner ring at $\simeq 0.5$ au and end up at 0.7-1 au, Type-I
  migration would have to be directed outward, for example because magnetically driven winds reduced the
  surface gas density in the inner part of the disk. Mercury was presumably left behind near the original inner
  ring location.  
\item With the total initial mass of 2.1 $M_{\rm Earth}$, the best results were obtained with 1.4 $M_{\rm Earth}$
  in the inner ring and 0.7 $M_{\rm Earth}$ in the inner ring. The Earth grows by first accreting planetesimals
  from the reduced inner-ring material and later collects $\sim 30$\% of more oxidized material from the
  outer source. This helps to explain the chemical composition of the Earth mantle (Rubie et al. 2011, Dale
  et al. 2025).
\item The two source reservoir model, with Earth growing from the inner ring and Mars accreting from the outer
  source, can potentially explain the isotopic differences between the Earth and Mars. It suggests that the
  Moon-forming impactor, Theia, could have formed from materials similar in isotopic composition to that of
  the Earth.
\item We found that the implantation efficiency of inner ring planetesimals to the asteroid belt is excessively
  low ($<3\times 10^{-7}$). This means that the inner ring reservoir should be unsampled in the current meteorite
  collection, which is consistent with the inferred properties of inner ring planetesimals (Dale et al. 2025;
  similar to Enstatite Chondrites in oxidation but enriched in refractories and s-isotope products).
\item We often register large impacts on Mars at $\sim 200$ Myr after $t_0$. This could provide a justification
  to the suggestion of Morbidelli et al. (2018) that the Mars crater record requires a global resurfacing event,
  probably related to the formation of the Borealis basin (Marinova et al. 2008, Nimmo et al. 2008), at
  $\sim 4.4$ Gyr ago.  
\item Some of the parameters investigated in this work do not seem to have much influence on the final results.
  For example, the initial size of planetesimals, which determines the strength of aerodynamic drag, does
  not significantly influence the results. The results are more fundamentally affected by the later stages
  when protoplanets form, migrate and experience giant impacts.
\item We found that the results with perfect mergers, an assumption adopted in most of our simulations, were not
  fundamentally changed when our additional simulations accounted for the hit-and-run and disruptive collisions. 
  In the simulations with collisional fragmentation Mercury's terminal mass was $\sim 30$\% lower then in the 
  simulations with perfect mergers. The mantle removal during collisions can help to increase Mercury's CMF. 
\item The effects of stochastic forces from gas disk turbulence will need to be investigated in more detail.
  Here we tentatively found that a modest level of stochastic forcing from turbulence can slightly improve the
  results for Mercury and Earth/Venus separation, but we did not investigate this issue in enough detail for the
  results to be conclusive. 
\end{enumerate}   
\acknowledgements

\begin{center}
{\bf Acknowledgments} 
\end{center} 
\vspace*{-3.mm}
The simulations were performed on the NASA Pleiades Supercomputer. We thank the NASA NAS computing division for continued
support. We thank the Center for Lunar Origin and Evolution (CLOE), a team in NASA's SSERVI program, for their support
for this work (cooperative agreement 80NSSC23M0176). The work of D.N., W.F.B. and R.D. was funded by this program.
A. M. and M.G. acknowledge support from the ERC grant N. 101019380 (HolyEarth). We thank Matt Clement for very helpful 
comments on the submitted manuscript.

\clearpage
\begin{table}
\centering
{
\begin{tabular}{lrrrrl}
\hline \hline
 \\   
 model                & V/E           & Mars       & Merc.       & $\Delta a$ & Notes\\   
                      & \%            & \%         & \%            & \%         & \\    
\hline
\multicolumn{6}{c}{\it Annulus/Ring models, no gas disk}\\
{\tt hansen2}         & 65            & 35         &  8.6          &  5.7       & {\footnotesize 1000 sims for Hansen (2009)}\\ 
{\tt ring4}           & 62            & 37         &  7.2          &  5.3       & {\footnotesize ring model, $r_1=0.85$ au}\\ 
{\tt ring401}         & 53            & 37         &  15           &  4.0       & {\footnotesize {\tt ring4} with 1000 planetesimals}\\ 
{\tt \bf ring402}     & {\bf 63}      & {\bf 38}   &  {\bf 7.7}    &  {\bf 10}  & {\footnotesize $r_1=0.85$ au, $\sigma_1=0.03$ au}\\ 
{\tt ring403}         & 49            & 32         &  3.9          &  4.4       & {\footnotesize $r_1=1$ au, $\sigma_1=0.1$ au}\\ 
\multicolumn{6}{c}{\it Ring models and background, no gas disk}\\
{\tt back10}           & 59            & 38         &  30           &  5.9       & {\footnotesize ring with 10\% background}\\ 
{\tt \bf back20}     & {\bf 55}      & {\bf 37}   &  {\bf 47}     &  {\bf 7.4} & {\footnotesize ring with 20\% background}\\ 
{\tt back30}         & 49            & 35         &  45           &  4.1       & {\footnotesize ring with 30\% background}\\ 
{\tt 0.01au}          & 56            & 41         &  47           &  7.2       & {\footnotesize narrow ring, $\sigma_1=0.01$ au}\\ 
{\tt nes21}           & 62            & 17         &  40           &  9         & {\footnotesize Nesvorn\'y et al. (2021), 100 sims}\\ 
{\tt mmsn}            & 26            & 11         &  0            &  0         & {\footnotesize gas disk with $\alpha_0=-1$} \\ 
\multicolumn{6}{c}{\it Ring models with convergent migration}\\
{\tt woo24}           & 23            & 9          &  26           &  0         & {\footnotesize Woo et al. (2024), 300 sims}\\ 
{\tt gas1}            & 39            & 15         &  36           &  14        & {\footnotesize {\tt back20} with convergent migration}\\ 
{\tt ring05}          & 39            & 7.7        &  31           &  44        & {\footnotesize $r_1=0.5$ au, stronger migration}\\ 
{\tt ring03}          & 27            & 1          &  56           &  45        & {\footnotesize $r_1=0.3$ au, stronger migration}\\ 
{\tt \bf back05}      & {\bf 49}      & {\bf 13}   &  {\bf 43}     &  {\bf 55}  & {\footnotesize $r_1=0.5$ au with 20\% background}\\ 
\multicolumn{6}{c}{\it Extended planetesimal disks with convergent migration}\\
{\tt power1}          & 49            & 12         &  24           &  11        & {\footnotesize power law $\gamma=-2$} \\ 
{\tt power2}          & 50            & 15         &  19           &  14        & {\footnotesize power law $\gamma=-2$, 5000 g cm$^{-2}$} \\ 
{\tt power3}          & 48            & 12         &  22           &  13        & {\footnotesize power law $\gamma=0$, 5000 g cm$^{-2}$} \\ 
{\tt hump1}           & 27            & 27         &  47           &  13        & {\footnotesize $\sigma_1=0.3$ au } \\ 
{\tt hump2}           & 37            & 15         &  43           &  4.5       & {\footnotesize $\sigma_1=0.35$ au, 3000 g cm$^{-2}$} \\ 
\hline \hline
\end{tabular}
}
\caption{The probability of tested formation models to satisfy different success criteria.
  The success criteria were described in Section 3.1. The best models are shown in bold.}
\end{table}

\clearpage
\begin{table}
\centering
{
\begin{tabular}{lrrrrl}
\hline \hline
 \\   
 model                & V/E           & Mars       & Merc.       & $\Delta a$ & Notes\\   
                      & \%            & \%         & \%            & \%         & \\    
\hline
\multicolumn{6}{c}{\it The two-source models}\\
{\tt \bf model203}        & {\bf 50 }  & {\bf 35}  &  {\bf 32}     &  {\bf 33} & {\footnotesize $r_1=0.6$ au, $r_2=1.7$ au, 3000 g cm$^{-2}$} \\ 
{\tt model211}        & 52            & 28         &  20           &  27       & {\footnotesize 1:1 split, $w_2=1/2$} \\ 
{\tt model212}        & 50            & 30         &  35           &  37       & {\footnotesize 3:1 split, $w_2=1/4$} \\ 
{\tt model213}        & 46            & 32         &  23           &  16       & {\footnotesize 1500 g cm$^{-2}$} \\ 
{\tt model214}        & 50            & 20         &  36           &  47       & {\footnotesize 6000 g cm$^{-2}$} \\ 
\multicolumn{6}{c}{\it Inner ring location and width}\\
{\tt model218}        & 48            & 30         &  22           &  35       & {\footnotesize $r_1=0.7$ au} \\ 
{\tt model217}        & 51            & 29         &  40           &  31       & {\footnotesize $r_1=0.5$ au} \\ 
{\tt model222}        & 46            & 34         &  58           &  27       & {\footnotesize $r_1=0.4$ au}\\ 
{\tt model223}        & 43            & 33         &  65           &  34       & {\footnotesize $r_1=0.3$ au, Mercury too close}\\ %
{\tt model225}        & 48            & 34         &  36           &  28       & {\footnotesize $r_1=0.5$ au, $\sigma_1=0.025$ au}\\ 
{\tt model224}        & 43            & 31         &  44           &  18       & {\footnotesize $r_1=0.5$ au, $\sigma_1=0.1$ au}\\ 
{\tt model235}        & 38            & 38         &  15           &  0        & {\footnotesize $r_1=0.5$ au, $\sigma_1=0.2$ au}\\ 
\multicolumn{6}{c}{\it Outer source location and width, $r_1=0.5$ au}\\
{\tt model216}        & 51            & 21         &  26           &  42       & {\footnotesize $r_2=1.5$ au} \\ 
{\tt model215}        & 44            & 38         &  32           &  20       & {\footnotesize $r_2=1.9$ au} \\ 
{\tt model229}        & 40            & 37         &  40           &  29       & {\footnotesize $r_2=2.1$ au}\\
{\tt model219}        & 48            & 30         &  31           &  38       & {\footnotesize $r_2=1.7$ au, $\sigma_2=0.2$ au} \\ 
{\tt model227}        & 42            & 35         &  37           &  38       & {\footnotesize $r_2=1.7$ au, $\sigma_2=0.3$ au}\\
{\tt model234}        & 48            & 39         &  42           &  31       & {\footnotesize $r_2=1.7$ au, $\sigma_2=0.4$ au}\\%
{\tt model228}        & 44            & 37         &  39           &  22       & {\footnotesize $r_2=1.9$ au, $\sigma_2=0.3$ au}\\
\multicolumn{6}{c}{\it Auxiliary models}\\
{\tt model231}        & 47            & 32         &  32           &  27       & {\footnotesize $N=1000$}\\ 
{\tt model220}        & 49            & 36         &  16           &  21       & {\footnotesize $\kappa=0$}\\ 
{\tt model221}        & 51            & 27         &  45           &  38       & {\footnotesize $r_1=0.5$ au, $r_2=1.9$ au, 6000 g cm$^{-2}$}\\ %
{\tt model233}        & 31            & 36         &  32           &  41       & {\footnotesize model203 with fragmentation}\\ %
\hline \hline
\end{tabular}
}
\caption{The probability of two-source models to satisfy different success criteria.
  The success criteria were described in Section 3.1. Our reference {\tt model203}, shown in bold, has the following parameters:
  $M_{\rm tot}=2.1$ $M_{\rm Earth}$, $r_1=0.6$ au, $\sigma_1=0.05$ au, $r_2=1.7$ au, $\sigma_2=0.1$ au, $\Sigma_0=3000$ g
  cm$^{-2}$, $\tau=1$ Myr, $\kappa=3 \times 10^{-6}$, 2:1 mass split between the inner and outer sources ($m_1=1.4$
  $M_{\rm Earth}$, $m_2=0.7$ $M_{\rm Earth}$; $w_2=1/3$), and $w_3=0$. The other two-source models adopted the
  same parameter values except for the ones noted in the last column.}
\end{table}

\clearpage

\clearpage

\clearpage

\clearpage

\clearpage

\clearpage

\clearpage

\clearpage

\clearpage

\clearpage

\clearpage

\clearpage

\clearpage

\clearpage

\clearpage

\clearpage

\clearpage

\end{document}